\documentclass[aps,superscriptaddress,nofootinbib]{revtex4}

\usepackage{graphicx}
\usepackage{amssymb,amsmath}
\newcommand{\nin}{\noindent}
\newcommand{\be}{\begin{equation}}
\newcommand{\ee}{\end{equation}}
\newcommand{\bea}{\begin{eqnarray}}
\newcommand{\eea}{\end{eqnarray}}
\newcommand{\nn}{\nonumber\\}

\newcommand{\miso}{\frac{1}{2}}
\newcommand{\alp}{\alpha '}

\newcommand{\da}{\dot{a}}
\newcommand{\dda}{\ddot{a}}
\newcommand{\ddda}{\dddot{a}}

\begin{document}

\begin{flushleft}
KCL-PH-TH/2012-27 \\
LCTS/2012-30
\end{flushleft}

\title{Stringy Models of
  Modified Gravity: Space-time defects and Structure Formation}

\vspace{1cm}
\author{Nick E. Mavromatos}
\affiliation{King's College London, Department of Physics, Strand,
London WC2R 2LS, UK.}
\affiliation{CERN, Theory Division, CH-1211  Geneva 23, Switzerland}

\author{Mairi Sakellariadou}
\affiliation{King's College London, Department of Physics, Strand,
London WC2R 2LS, UK.}

\author{Muhammad Furqaan Yusaf }
\affiliation{King's College London, Department of Physics, Strand,
London WC2R 2LS, UK.}

\begin{abstract}

Starting from microscopic models of space-time foam, based on brane
universes propagating in bulk space-times populated by D0-brane
defects (``D-particles''), we arrive at effective actions used by a
low-energy observer on the brane world to describe his/her
observations of the Universe. These actions include, apart from the
metric tensor field, also scalar (dilaton) and vector fields, the
latter describing the interactions of low-energy matter on the brane
world with the recoiling point-like space-time defect
(D-particle). The vector field is proportional to the recoil velocity
of the D-particle and as such it satisfies a certain constraint. The
vector breaks locally Lorentz invariance, which however is assumed to
be conserved on average in a space-time foam situation, involving the
interaction of matter with populations of D-particle defects. In this
paper we clarify the role of fluctuations of the vector field on
structure formation and galactic growth. In particular we demonstrate
that, already at the end of the radiation era, the (constrained)
vector field associated with the recoil of the defects provides the
seeds for a growing mode in the evolution of the Universe.  Such a
growing mode survives during the matter dominated era, provided the
variance of the D-particle recoil velocities on the brane is larger
than a critical value. We note that in this model, as a result of
specific properties of D-brane dynamics in the bulk, there is no issue
of overclosing the brane Universe for large defect densities. Thus, in
these models, the presence of defects may be associated with
large-structure formation.  Although our string inspired models do
have (conventional, from a particle physics point of view) dark matter
components, nevertheless it is interesting that the role of
``extra'' dark matter is also provided by the population of massive
defects.  This is consistent with the weakly interacting character of
the D-particle defects, which predominantly interact only
gravitationally.

\end{abstract}

\maketitle

\section{Introduction and Motivation}
Relativistic modified gravity theories (\emph{e.g.} models in
Refs.~\cite{bekenstein,moffat}) have been presented originally as
field theoretic alternatives to dark matter models, offering support
to the heuristic Modified Newtonian Dynamics
approach~\cite{mond}. Although, at least in their simplest form, such
models may be ruled out by means of recent precision astrophysical
measurements using gravitational lensing~\cite{lensing}, nevertheless
some of their features may characterise microscopic theories of
quantum gravity, in harmonic co-existence with substantial dark matter
components, thereby avoiding the above-mentioned stringent
constraints.

For instance, it has been shown in Ref.~\cite{Mavromatos:2007sp}, that
the bi-metric nature of the TeVeS (Tensor-Vector-Scalar) model, as
well as the existence of a Lorentz-violating vector field $A_\mu$, do
characterise the low-energy limit of certain string-theory models of
space-time foam~\cite{foam,foam2}, involving three-brane
(``D3-brane'', a three-space-dimensional Dirichlet brane)
worlds~\cite{Polchinski,johnson} embedded in higher-dimensional (bulk)
space-times punctured by D0-brane (``D-particle'') defects. In such
models, the D3-brane representing our Universe, which may be obtained
from appropriate compactification of a higher-dimensional brane, moves
in the bulk and in this way the D-particles cross it, resembling ---
from a D3-brane observer viewpoint --- flashing ``on'' and ``off''
foamy structures (``D-foam''). There are topologically non-trivial
interactions of open string states, attached on the D3-brane and
representing ordinary matter, with such defects, involving splitting
of the initial string and creation of intermediate string states,
stretched between the D-particle and the D3-brane universe, exhibiting
length oscillations. Such processes result in local distortions of the
neighbouring space-time~\cite{foam}
 \begin{equation}\label{recoilmetric}
 g_{\mu\nu} = \eta_{\mu\nu} + h_{\mu\nu}~, \quad h_{0i} = v_i = g_{\rm
 s} \frac{\Delta k_i}{M_{\rm s}}~,
 \end{equation}
to leading order in the recoil velocities $v_i \ll 1$, due to recoil
of the D-particle defect to conserve energy and momentum during the
scattering.  In the above formula, $g_{\rm s}$ is the (weak) string
coupling, $M_{\rm s}$ is the string mass scale, which is in general
different from the four-dimensional Planck scale $M_{\rm P} =
1.2. \times 10^{19}$~GeV, and $v_i $ is the recoil three-velocity of
the D-particle, with $\Delta k_i$ the momentum transfer of the open
string state.

Due to electric charge conservation, only electrically neutral matter
 interactions with the D-particles~\footnote{This is the case of type
  IIA string theory, which allows point-like branes. In the
  phenomenologically more realistic type IIB string models, there are
  no D0-brane configurations allowed. In such a case, one can still
  construct consistent D-foam models~\cite{li} in which the role of
  the D-particle is played by D3-branes wrapped up around small three
  cycles. In such a case, electrically charged excitations do interact
  non-trivially with the D-foam, but these interactions are
  significantly suppressed compared with those of the neutral
  excitations.} are allowed. In this sense, there is a naturally
induced \emph{bi-metric} structure in the model, given that it is only
the electrically neutral excitations of the low-energy effective
theory that predominantly feel the induced metric
Eq.~(\ref{recoilmetric}).  Moreover, like in TeVeS models, there exist
\emph{vector field} structures, provided by the recoil velocities of
the D-particles, which break Lorentz invariance locally.

As pointed out by the authors of Ref.~\cite{dodelson}, it is the
vector field of the TeVeS theory that they were working with, which
plays the crucial role in reproducing the observed power spectrum by
generating a cosmological instability that produces the large cosmic
structures of today's Universe. It is important to note that the
arguments on the role of the vector field in reproducing large scale
structures and the correct phenomenology seem to be generic, in the
sense that they do not depend on the detailed Lagrangian of the
original TeVeS theory proposed by Bekenstein~\cite{bekenstein}. More
precisely, it is only the basic important features of the theory,
namely the existence of the two metrics (bi-metric theory) and of the
(Lorentz-violating) ``aether-like'' vector field, that appear to be
important in this respect.

The basic aim of this paper is to extend the link between the above
mentioned features and the deduced cosmological perturbation theory,
in the framework of some specific backgrounds of string theory
proposed some time ago~\cite{foam} as consistent candidates for a
quantum space-time foam background in string theory. Specifically, we
shall construct a low-energy effective action, describing (partly
though) the low-energy aspects of such a stringy space-time foam, and
argue that it acquires the form of a modified gravity theory, with
scalar and vector fields arising naturally, as a result of the
interaction of matter with the D-particle space-time defects in the
foam.  Then, we shall consider the equations for cosmological
perturbations, solve them to linear order for small perturbations, and
argue that the vector fields in the theory play a non-negligible
role in structure formation, in the spirit of
Ref.~\cite{dodelson}. However, we stress that such theories do not
provide alternative to dark matter, given that the latter exists
naturally in the superstring-inspired models we are working with,
represented by stable superpartners to the standard model matter that
characterises the low-energy excitations of the models.  In this
sense, our approach is entirely different from models which are
alternatives to dark
matter~\cite{bekenstein,moffat}. Nevertheless, the D-particle
defects themselves play a role analogous to dark matter, given that
their recoil furnishes the Universe with an effective vector field,
whose perturbations in late eras of the Universe (radiation and matter
dominated) do posses growing modes and can participate non-trivially
in large structure formation.  Thus the massive D-particles can, in
addition to baryonic and other type of matter, contribute to galaxy
and galactic cluster formation.

To set-up the framework we shall use, we briefly review in
Section~\ref{II} the brane D-foam model that will serve as our
microscopic framework for discussing modified gravity in the low
energy limit. First we present the underlying formalism of the
world-sheet deformation that describes the interaction of an open
string state with a recoiling D-particle and derive the associated
metric distortion of the neighbouring space-time, due to the recoil of
the defect.  We explain in detail the emergence of a vector ``gauge''
field $A_\mu$ as a result of this interaction, which will play a
crucial role in our analysis in providing us with the growing mode
that can lead to structure formation. The vector fields are
proportional to the recoil velocities of the D-particles, and as such
they satisfy a certain type of constraint
(\emph{cf}. Eq.~(\ref{constraint})), which is crucial for the
appearance of the growing mode.  We discuss in Section~\ref{III} more
general background configurations for the dilaton, graviton and gauge
fields, in a way consistent with the conformal invariance of the
world-sheet of the string. We first present the Dirac-Born-Infeld (DBI)
action on 3-brane worlds and we then proceed with space-time curvature
correction terms in the DBI action. In Section~\ref{secac1} we study
the four-dimensional induced effective action on a D3-brane world for
the gravitational, dilaton and gauge fields. This is the action that
we subsequently use to analyse the role of vector perturbations in
structure formation. In Section~\ref{bkcgd} we derive the equations of
motion for the vector, metric and dilaton fields and consider their
cosmological perturbations.  In Section \ref{sec:growth} we discuss
the appearance of a growing mode in both the radiation and matter
dominated eras of the brane Universe, provided a sufficiently large
variance of recoil velocities of the D-particles exists. In
this sense, the defects act as a dark matter component, which is
in addition to the conventional (from a particle physics viewpoint) dark
matter components of string effective models, e.g. supersymmetric
partners of matter or graviton fields.  Finally in Section
\ref{sec:concl} we give our conclusions and outlook.

We use the following conventions: for the metric we use the signature
$(-, +, +, +)$ and the Riemann curvature tensor is defined as
$R^\alpha _{\beta \, \gamma \, \delta} = \partial_\delta \,
\Gamma^\alpha_{\,\, \beta\, \gamma} + \Gamma^\lambda_{\,\, \beta \,
\gamma} \, \Gamma^\alpha_{\,\, \lambda \, \delta} - ( \gamma
\leftarrow\rightarrow \delta) $.

\section{Review of D-foam Inspired Modified gravity Effective Theories}
\label{II}

\begin{figure}[t]
\centering \includegraphics[width=5.5cm]{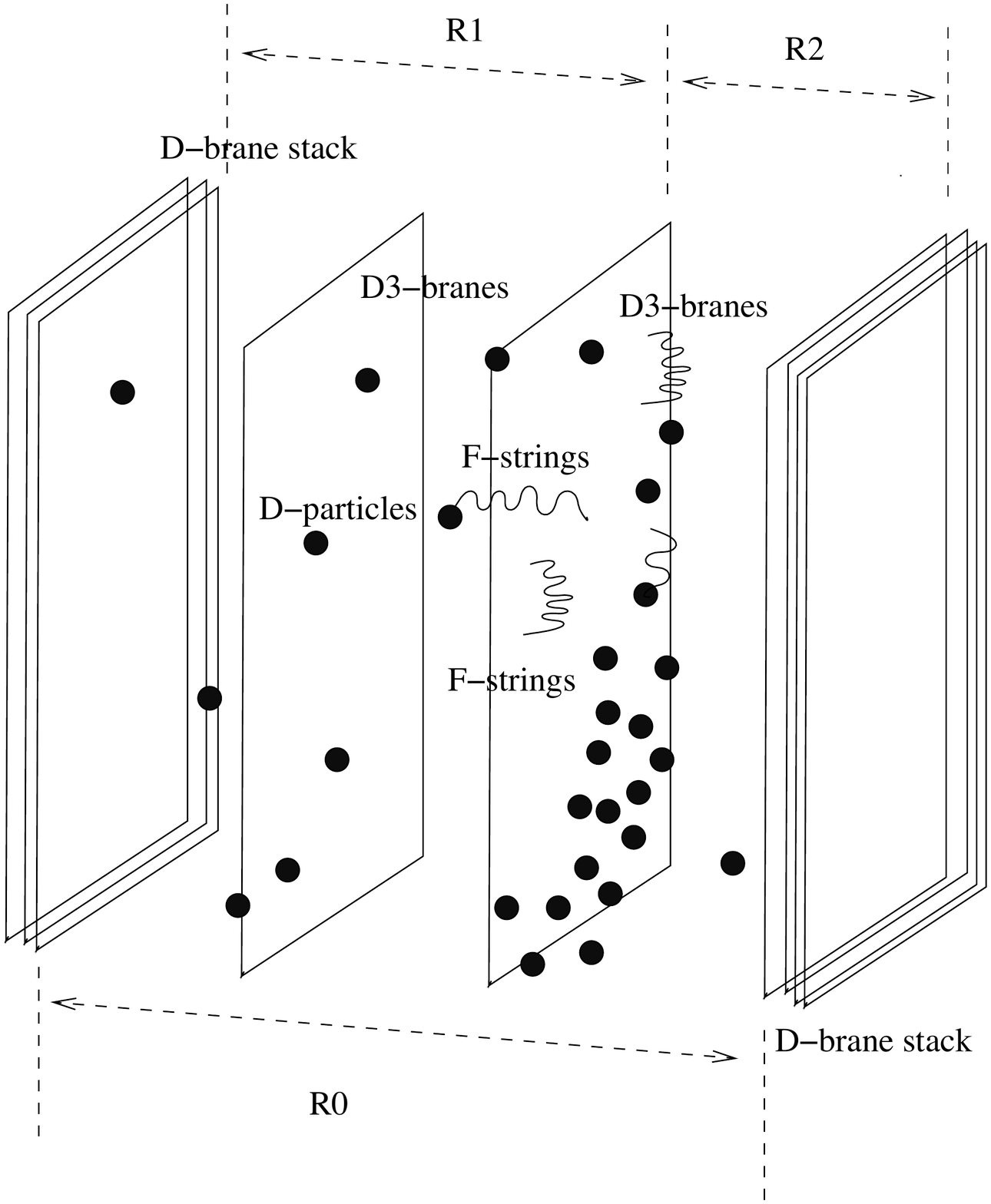} \caption{Schematic
representation of a generic D-particle space-time foam model.  The
model of Ref.~\cite{foam}, which acts as a prototype of a D-foam,
involves two stacks of D8-branes, each stack being attached to an
orientifold plane.}

\label{fig:recoil}
\end{figure}

The prototype model that we shall use as a microscopic motivation of
our considerations in this work is the so-called D-brane model of
space-time ``foam'', in which our Universe is represented as a
(compactified) three-brane world moving in a higher-dimensional bulk
space-time which is populated by point-like D0-brane defects
(D-particles).  Such a model is appropriate for type IIA string theory
and the associated supergravity effective field theory in the bulk and
was first presented in Ref.~\cite{foam}.  There is a relative motion
of the defects with respect to the branes, which implies that, as the
brane universe moves in the bulk, the D-particles cross it, so that
from the point of view of an observer on the brane they look like
``flashing on and off'' foamy structures (hence the terminology
``D-particle foam'' model).  In string perturbation theory, it can be
shown that for an adiabatic relative motion between a D-particle
defect and the brane, with the defect moving perpendicularly to the
brane, the induced potential on the brane, due to open strings
stretched between the D-particle and the brane, depends on both the
relative velocity $v_\perp$ and the distance of the bulk defect from
the brane:
\begin{eqnarray}
\mathcal{V}_{D0-D8}^{\rm long}=+\frac{r\,(v_{\perp}^{{\rm
long}})^{2}}{8\pi\alpha^{\prime}}~,~r\gg\sqrt{\alpha^{\prime}}~
\label{long-1}
\end{eqnarray}
However, a D-particle close to the D3-brane (compactified D8), at a
distance $r'\ll\sqrt{\alpha'}$, moving adiabatically in the
perpendicular direction with a velocity $v_{\perp}^{{\rm short}}$,will
induce the potential :
\begin{equation}\label{pot1} \mathcal{V}_{D0-D8}^{\rm short}=
- \frac{\pi\alpha^\prime (v_\perp^{\rm short})^2}{12{r'}^3}.
\end{equation}
This difference in sign, implies that, one can arrange for the
 densities of far away and nearby bulk D-particles, which are not in
 general homogeneous, to be such that the total contribution to the
 brane world's vacuum energy is always sub-critical, so that issues
 such as \emph{over-closure} of the Universe by a significant
 population of D-particle defects can be \emph{avoided}~\cite{foam2}.
 Consequently, the astrophysical data can only constrain the total
 energy density on the brane due to the defects, that is the algebraic
 sum of the above two contributions in Eqs.~(\ref{long-1}),
 (\ref{pot1}). Hence, the density of defects on the brane remains
 unconstrained. We shall come back to this point in Section
 \ref{sec:growth}.

These considerations imply that, at least at early eras of the
Universe, we can plausibly assume that the brane Universe passes
through bulk regions which are densely populated by D-particles, in
such a way that there is a sufficiently high density of D-particles on
the brane, such that the particle excitations, represented by open
strings with their ends attached on the brane, propagate in a
``\emph{medium}'' of such D-particle defects. Topologically
non-trivial interactions between the defects and matter strings can
occur, involving the capture and splitting of open strings by the
defects and creation of stretched strings between the defects and the
brane during such processes~\cite{foam,foam2} (\emph{cf.}
Fig.~\ref{fig:recoil}). The capture process is represented from a
world-sheet point of view as an impulse which leads to a metric
deformation of the neighbouring space-time due to the recoil of the
D-particle. The induced metric distortions depend on both the
space-time coordinates and the momentum transfer of the matter
particle during its scattering with the defect. The recoil of the
defect breaks locally Lorentz invariance of the space-time as a result
of its momentum direction, and in the effective low-energy limit is
described as a vector field $A_\mu$, which should not be confused with
the electromagnetic field. On average, over large populations of
defects Lorentz invariance may be restored in the sense of having a
zero vacuum expectation value of the vector field, but non-zero
variances. The purpose of this work is to examine the effective
low-energy target space action describing the interaction of matter
excitations with the D-particles, through the field $A_\mu$ and
discuss in this context the potential role of the perturbations of
the vector field in inducing a growing mode that could participate in
large-scale structure formation at redshifts of order $z \sim 1$. As a
result of charge conservation, charged open string excitations cannot
interact with the (neutral) defects via the above-mentioned
splitting. Only electrically neutral excitations are therefore
interacting predominantly with the D-particle foam. Cosmologically it
is therefore the neutrinos and photons that feel mostly the effects of
the foam at these early eras, and hence it is these particles that can
find themselves propagating in the background of recoiling D0-branes,
and hence of the associated vector field $A_\mu$.  We shall show here
that D-particles can play the role of a dark matter component, in
addition to either supersymmetric partners of standard model
excitations or neutrinos.

We commence our discussion towards a formulation of an effective
low-energy action for this type of stringy interactions, including the
dynamics of the vector field itself, by first reviewing briefly the
underlying formalism of the world-sheet deformation that describes the
(impulse) interaction of an open string state with a recoiling
D-particle~\cite{kogan}.  The pertinent vertex operator on the
world-sheet boundary $\partial\Sigma$, within the world-sheet
$\sigma$-model reads:
\be \label{recoil} \mathcal{V}_{\rm{recoil}}^{\rm impulse} =
\int_{\partial\Sigma} u_i X^0 \Theta_\epsilon (X^0) \partial_n X^i~,
\ee
where $X^0$ is the target-space time coordinate, obeying Neumann
boundary conditions, and $X^i$ ($i=1,2,3$) are the spatial coordinates
on the brane, obeying Dirichlet boundary conditions. The notation
$\partial_n$ denotes normal derivatives on the world-sheet.  By $u_i$
we denote the recoil relativistic three-velocity of the D-particle
($u_i=\gamma v_i = \gamma g_{\rm s} \Delta k_i / M_{\rm s}$, with
$\gamma=(1-\vec v^2)^{-1/2}$ the corresponding Lorentz factor, and
$\Delta k_i$ the momentum transfer of the stringy-matter excitation).
The operator
\be \Theta_{\epsilon } (X^0) = -i \int_{-\infty}^\infty
\frac{d\omega}{\omega - i\epsilon} e^{i\omega X^0}~~\mbox{with}~~
\epsilon \to 0^+~, \ee
is a regularised operator. It can be shown~\cite{kogan} that for
$\epsilon^{-2} = 2{\rm ln}(A/a^2)$, where $A/a^2$ is the area of the
world-sheet in units of a world-sheet Ultra-Violet (UV) cut-off length
scale $a$, the vertex operator in Eq.~(\ref{recoil}) satisfies an
appropriate logarithmic conformal field theory.  The operator has an
anomalous dimension $-\epsilon^2/2 \to 0^+$ in the Infra-Red (IR) fixed
point $A/a^2 \to \infty$, where we shall work from now
on~\cite{Mavromatos:2007sp}.

The impulse operator in Eq.~(\ref{recoil}) is written in the co-moving
frame of the recoil D-particle. In the frame of an observer on the
D$3$-brane, the covariantised form of this expression should be
used instead~\cite{Mavromatos:2007sp}. We can write this as a total
world-sheet derivative in the following way:
\be\label{covariant} \int_{\Sigma}\partial_a\left( u_\mu u_\nu X^\nu
\Theta_\epsilon (u_\rho X^\rho) \partial^a
X^\mu\right)~~\mbox{with}~~\epsilon \to 0^+~, \ee
where $u_\mu$ is a four-velocity vector satisfying:
\be \label{u2} u_\mu u^\mu=-1~,  \ee
and $a=1, 2$ is a world-sheet index. Note that above, using the
Neumann boundary conditions, $\partial_n X^0=0$, for the temporal
coordinate $X^0$, thereby implying an ambiguity in Eq.~(\ref{recoil})
when written as a total world-sheet derivative, allowed us to write
the covariant form Eq.~(\ref{covariant}).

As discussed in detail in Ref.~\cite{Mavromatos:2007sp}, where we
refer the interested reader for details, the bulk operator
Eq.~(\ref{covariant}) is conformal on the world-sheet in the sense of
having conformal dimension two.  However, the target-space metric
deformation implied by the vertex Eq.~(\ref{covariant}) does not
connect smoothly with the flat Minkowski space-time before the capture
of the open string by the D-particle.  Indeed, by expanding the
derivative in Eq.~(\ref{covariant}) and ignoring terms of the form
$\partial_\alpha \partial^\alpha X^\mu$, that can be eliminated
on-shell in the world-sheet theory, we recognise a contribution
similar to the free sigma-model action propagating in a curved
background $g_{\mu\nu}$, namely $\int _{\Sigma} g_{\mu\nu}\partial_a
X^\mu \partial^a X^\nu$.  In this sense, there is a target-space
deformation due to the D-particle recoil:
\be\label{deform1} \delta g_{\mu\nu} = u_\mu u_\nu \Theta_{\epsilon
  \to 0^+} (u_\rho X^\rho).  \ee
To ensure a smooth connection with Minkowski space-time before the
collision/capture, one can couple to this deformed $\sigma$-model a
linear dilaton background term of the form~\cite{Mavromatos:2007sp}:
\bea\label{dilaton}
\int_\Sigma d^2\xi \phi R^{(2)}~,\quad \phi =u_\mu X^\mu ~, \eea
where $d^2\xi$ is the covariant measure over the curved world-sheet of
curvature $R^{(2)}$.  Coupling the vertex operator
Eq.~(\ref{covariant}) to $e^\phi$ factors does not affect the
conformal invariance, since the conformal dimension of $e^\phi$ in the
presence of the linear-dilaton background Eq.~(\ref{dilaton}) is
$u_\mu (u^\mu - u^\mu) = 0$. Hence the complete vertex operator for
recoil, compatible with conformal invariance at the IR fixed-point
$\epsilon \to 0^+$, turns out to be~\cite{Mavromatos:2007sp}:
\be \mathcal{V}_{\rm recoil}=\int_{\Sigma} e^\phi \partial_a\left(
u_\mu u_\nu X^\nu \Theta_{\epsilon \to 0^+} (u_\rho X^\rho) \partial^a
X^\mu\right)~.  \ee
Acting with the derivative in the above formula, we can see that the
deformed target-space metric has the form
\be\label{deform}
\delta g_{\mu\nu} = e^\phi u_\mu u_\nu \Theta_{\epsilon \to 0^+}
(u_\rho X^\rho).  \ee
However, to ensure a smooth connection with the flat metric at the
origin of the boosted time, $u_\rho X^\rho=0$, the above metric
deformation Eq.~(\ref{deform}) must also contain extra dilaton terms,
such that the induced metric is of the form:
\bea\label{finaldeform}
g_{\mu\nu}^{\rm neutral~matter} &=& \eta_{\mu\nu} - \left(e^\phi
-e^{-\phi}\right)u_\mu u_\nu~,\nonumber\\
\phi &=& u_\mu X^\mu~.  \eea
Note that the $e^{\phi}$ correction corresponds to an operator on the
world-sheet of the string with \emph{zero} conformal dimension, $
-u_\mu ( - u^\mu - u^\mu ) + 2 = 2 (u_\mu u^\mu + 1) =0~,$ due to
Eq.~(\ref{u2}).  This latter deformation leads to \emph{departure from
  criticality} of the associated $\sigma$-model and thus Liouville
dressing is required~\cite{ddk}.

To this end~\cite{Mavromatos:2007sp}, we first notice that the linear
dilaton implies a sub-critical string with $Q^2 = u_\mu u^\mu = -1 <
0$.  This can become conformal if one uses a {\it space-like}
Liouville mode~\cite{ddk} $\rho$ to ``dress'' the above-mentioned
metric deformations by multiplication with exponential operators
$e^{\alpha_i \rho}$ (with $i=1,2$); $\alpha_i$ denote the Liouville
``anomalous'' dimensions.  In the presence of the world-sheet
background charges in the $(\rho, X^\mu)$ extended target space time
of the Liouville-dressed world-sheet theory, induce world-sheet
curvature terms of the form
$$\int_\Sigma d^2\xi (u_\mu X^\mu + \rho)R^{(2)}$$
in the $\sigma$-model action. The conformal dimension of these
operators is $\alpha_i (\alpha_i + 1)$ for each $i=1,2$.  Restoration
of conformal invariance requires that the total conformal dimension of
the Liouville-dressed deformations is (1,1) in the (holomorphic,
anti-holomorphic) world-sheet sectors. It is straightforward then to
observe that the following dressed operators are conformal, amounting
to the choices $\alpha_i = \pm |Q| = \pm 1$ in the respective
Liouville anomalous dimensions,
\begin{equation}
\mathcal{V}_{({\rm dressed})}^{\lambda\nu} u_\lambda u_\nu =
\left(e^{-u_\mu X^\mu -\rho} - e^{u_\mu X^\mu + \rho}\right)\partial
X^\lambda {\overline \partial}X^\nu u_\lambda u_\nu~.
\end{equation}
These imply a dressed target-space-time metric in the extended
space-time $(\rho, X)$ of the form:
\begin{eqnarray}\label{finaldeform2}
g_{\mu\nu}^{\rm neutral~matter,~ dressed} (\rho, X) &=& \eta_{\mu\nu}
+ \left(e^{\Phi(\rho, X)} -e^{-\Phi(\rho, X)}\right)u_\mu u_\nu~,
\nonumber \\ g_{\rho \mu} &=& 0~,\nonumber\\
g_{\rho\rho}&=&+1~,\nonumber\\ \Phi (\rho, X) &=&- u_\mu X^\mu -
\rho~.
\end{eqnarray}
The extra space-like Liouville mode may thus be given the physical
interpretation of a bulk spatial dimension, in which recoil of the
defects does not take place. Here, our brane space-time is located at,
say, $\rho = 0$.  Note that this is just one example of a consistent
conformal theory.  In general, one may consider non-trivial
$\sigma$-model metrics $G_{\mu\nu}$, in which case the associated
dilatons will have a more complicated space-time dependence. In this
respect, one can discuss the effects of D-foam recoil in realistic
Friedman-Lema\^{i}tre-Robertson-Walker (FLRW) backgrounds, which
constitutes the subject of the present paper.

With the above considerations in mind, we take from now on the
following form of the deformed target-space metric:
\bea\label{metric} g_{\mu\nu}^{\rm neutral~matter}&=&G_{\mu\nu} +
(e^\phi-e^{-\phi}) u_\mu u_\nu~~~\mbox{with}~~~ \mu, \nu = 0, \dots
4~,\nonumber\\ \phi &=&\Phi(0, X^\mu)~~~\mbox{with}~~~ \mu, \nu = 0,
\dots 4~, \eea
due to D-particle recoil during their interaction with neutral matter.
Thus, we assume a three(spatial)- dimensional (D3-)brane universe from
now on for brevity (the latter may be obtained by appropriate
compactification of higher dimensional branes), ignoring bulk physics
as far as the recoil of D-particles is concerned, which we assume to be
dominant only on the brane world.

Finally, we remark that the recoil velocity field $u_\mu$ is elevated
to a dynamical one in this approach, as being part of a gauge
background field in the D-particle recoil, which obeys non-trivial
dynamics~\cite{Mavromatos:2007sp}. This can be seen in two equivalent
ways. In the first one, we use the fact that, upon $T$-duality (which
exchanges Neumann and Dirichlet boundary conditions and is assumed to
be an exact symmetry of the underlying string theory), the
(non-covariantised) vertex operator of recoil Eq.~(\ref{recoil}) is
related to the background gauge potential deformation~\cite{kogan}:
\be A_i = u_i X^0 \Theta (X^0)~,
\label{gauge}
\ee
assuming~\cite{kogan}, without loss of generality, a time axial gauge
$A_0 = 0$. This background corresponds for $X^0 > 0$ (after the
impulse) to a constant ``electric''-type field $E_i \equiv F_{0i} =
u_i$, where $F_{\mu\nu} = \partial_\mu A_\nu$ is the Maxwell field
strength of the gauge field.  This vector field describes the average
effects of the interaction of neutral matter with the background of
the recoiling D-particle and should not be confused with the
electromagnetic field of the standard model; it is a new degree of
freedom associated with the back reaction of the D-matter onto space
time.

Alternatively, without invoking T-duality, one may use the world-sheet
version of Stokes' theorem, to write down the boundary recoil
deformation Eq.~(\ref{recoil}) (before coupling to the
dilaton/Liouville) as a bulk deformation~\cite{mavroreview}:
\begin{eqnarray}\label{stokes}
\mathcal{V}_{\rm{recoil}}^{\rm impulse}&=&\frac{1}{2\pi\alpha '}
\int_{\Sigma}d^{2}\xi\,\varepsilon_{\alpha\beta} \partial^\beta \left(
\left[ u_{i}X^{0}\right] \Theta_\epsilon \left( X^{0}\right)
\partial^{\alpha}X^{i}\right) \nonumber \\ &=& \frac{1}{4\pi\alpha
'}\int_{\Sigma}d^{2}\xi\, (2u_{i})\,\varepsilon_{\alpha\beta}
\partial^{\beta }X^{0} \Bigg[\Theta_\epsilon \left(X^{0}\right) + X^0
\delta_\epsilon \left( X^{0}\right) \Bigg] \partial ^{\alpha}X^{i}~,
\end{eqnarray}
where $\varepsilon_{\alpha\beta}$ is the world-sheet Levi-Civita
antisymmetric symbol and $\delta_\epsilon (X^0)$ is an
$\epsilon$-regularised $\delta$-function. For relatively large times
after the impact at $X^0=0$ (which we assume here for our
phenomenological purposes), this is equivalent to a deformation
describing an open string propagating in an antisymmetric
$B_{\mu\nu}$-background corresponding to an external constant in
target-space ``electric'' field,
\begin{equation}
B_{0i} = F_{0i} \sim u_i ~~\mbox{and}~~ B_{ij}=0~~\mbox{for}~~ X^0 >
0~,
\label{constelectric}
\end{equation}
where the $X^0\delta (X^0)$ terms in the argument of the electric
field yield vanishing contributions in the large time limit, and hence
are ignored from now on. In this approach, we note that the presence
of the B-field leads to mixed-type boundary conditions for open
strings on the boundary $\partial \Sigma $ of world-sheet surfaces
with the topology of a disc; we consider here for concreteness and
relevance to our discussion below, that:
\begin{equation}
      g_{\mu\nu}\partial_n X^\nu + B_{\mu\nu}\partial_\tau X^\nu
      |_{\partial \Sigma} = 0~,
\label{bc}
\end{equation}
where $g_{\mu\nu} $ denotes the metric of target space-time. Formally,
the operator Eq.~(\ref{stokes}) is conformal on the world-sheet, in
the limit $\epsilon \to 0^+$~\footnote{Notice that to express the
gauge deformation in the frame of a D3-observer, it is not sufficient
to replace $X^0 \rightarrow u_\mu X^\mu $, in the argument of the
$\Theta_{\epsilon \to 0^+}(X^0)$, but also to transform the recoil
three-velocity vector $u_i$. However, for our purposes this is not
necessary.} in flat space-times. The generalisation to curved space,
and in particular to FLRW backgrounds, will be discussed in subsequent
sections.

This conformal gauge field background is assumed to exist together
with the metric and dilaton backgrounds, discussed previously.  The
dynamics of the gauge field describes physics in the so-called open
string channel, whilst the dilaton and metric backgrounds describe the
effects of the recoiling D-particles in the closed string channel.
The above considerations summarise therefore the features of the
space-time foam model which motivate the inclusion of vector and
scalar fields together with gravitational tensor fields.

The reader should notice a formal similarity of the metric
Eq.~(\ref{metric}) with the corresponding one in TeVeS
models~\cite{bekenstein}, including, in addition to the graviton
tensor field $g_{\mu\nu}$, also scalar (dilaton $\phi$) and vector
(recoil velocity-related gauge $A_\mu$) fields. However, in our case,
the interpretation is entirely different from the TeVeS models, and
moreover, given that we are dealing with ordinary (from a
phenomenological point of view) supersymmetric strings, there are
natural candidates for dark matter in our models (the lightest
(stable) supersymmetric partners to the standard model excitations),
co-existing with the scalar and vector structures. Thus, our models
are not meant to provide alternative to dark matter
scenarios. Nevertheless the existence of dilaton and vector fields can
affect certain cosmological features of the string universe, which
shall be explored in this article.

In a cosmological context, as already mentioned, one needs to
reconsider the conformal invariance conditions for metrics of FLRW
type that depend on the cosmic time. These conditions are equivalent
to equations of motion for the various target-space fields that are
derived from a low-energy string effective action.  We next proceed
therefore to discuss the effective low-energy action, which can
describe in general quantum fluctuations of the tensor, scalar and
vector backgrounds and in subsequent sections we shall discuss
background solutions to the respective equations of motion and
perturbations around them.  The background solution we shall deal with
in this work is characterised by a constant dilation but non trivial
FLRW metrics and recoil vector fields adapted to this geometry, which
will generalise the flat space solution (\ref{gauge}) mentioned above.

\section{Low-Energy Effective Lagrangian for D-foam models: a proposal}
\label{III}
Here, we shall generalise the arguments of the previous section to
discuss more general background configurations for the dilaton,
graviton and gauge fields, in a way consistent with the conformal
invariance of the world-sheet of the string.  We would like to search
for cosmological backgrounds to be applied in our cosmological tests
of the model. In particular, we will investigate the role of vector
fields to galaxy growth.  We shall construct first an effective
low-energy field theory for a graviton, dilaton and vector field
background in the context of the model described above and find
configurations for the background fields that are consistent with the
conformal invariance conditions of the $\sigma$-model theory,
generalising the analysis of Ref.~\cite{Mavromatos:2007sp}. Then, we
shall analyse the resulting modifications of gravity in a cosmological
context, in particular we will study their effects in galactic growth
and structure formation. The requirement of conformal invariance of
the string theory on the world-sheet is equivalent to the requirement
of the background fields satisfying equations of motion arising from
the target space-time effective action. As the vector-field part of
the action used to describe open strings ending on D-branes is
well-known and can be expressed in a closed form to all orders in the
string Regge slope $\alpha '$ (DBI action)~\cite{tseytlin}, we are
able to start from it and then study its modifications to include the
description of gravity in the bulk and on the brane.

\subsection{Dirac-Born-Infeld action on the 3-brane worlds }
We commence our analysis from the gauge field effective Lagrangian in
four-dimensional flat space-time, representing the case of propagating
open strings in gauge-field backgrounds, with their ends attached on a
D3-brane. The gauge field backgrounds and the open strings are not
propagating in the bulk. In the open string channel, it is known that
such a Lagrangian can be obtained to all orders in the $\alpha '$
expansion, and has the DBI form~\cite{tseytlin}:
\be \mathcal{L}_{\rm DBI} = -\frac{1}{g_{\rm s}} T_3 \sqrt{{\rm
    det}_4\left(\eta_{\mu\nu} + 2\pi \alpha ' F_{\mu\nu} \right)}~,
  \label{bi}
  \ee
where $T_3$ is the three-dimensional brane tension and $F_{\mu\nu}(A)
$ is the (Abelian) field strength.  It can be readily shown that a
constant electric field strength background, as is the case of the
recoil-velocity gauge field for times after the impulse, $X^0 > 0$, is
a consistent solution of the gauge field equations of motion derived
from this action, and thus a consistent (world-sheet conformal
invariant) configuration for the string.

In the presence of non-trivial dilatons, the string coupling $g_{\rm
  s}$ becomes $g_{{\rm s}0} e^\phi$ (where $g_{{\rm s}0} =e^{< \phi
  >}$ is a space-time constant). Moreover, in non-Minkowski
  backgrounds, the tensor $\eta_{\mu\nu}$ should be replaced by the
  background metric $g_{\mu\nu}$ and thus the DBI Lagrangian
  Eq.~(\ref{bi}) acquires the form:
\be \mathcal{L}_{\rm DBI} = -\frac{1}{g_{s0}}T_3 e^{-\phi} \sqrt{{\rm
    det}_4\left(g_{\mu\nu} + 2\pi \alpha ' F_{\mu\nu} \right)} +
    \dots~,
  \label{bicurved}
  \ee
where the $\dots$ include other terms, which are non-trivial in the
 case where
the field strength $F_{\mu\nu}$ is non-constant, and also include
curvature terms that give the graviton field dynamics. In the closed
string channel, of course, the latter terms are just the
Einstein-scalar curvature terms to lowest order in derivatives for the
gravitational sector. However, an interesting question arises as to
whether one may discover curvature terms coupled directly to the DBI
action, as a result of induced metrics in the open string channel.  It
is the point of this section to attempt to address such a question.

First of all we remark that the four-dimensional DBI action can be
expanded in derivatives, as appropriate for a low-energy
approximation, compared to the string scale $M_{\rm s}=1/\sqrt{\alpha
'}$, as follows~\cite{tseytlin}:
\be \label{dbi2} {\det } _4 \left(g_{\mu\nu} + 2\pi \alp F_{\mu\nu}
\right) = {\det } _4 g \left[1+ (2\pi\alp)^2 I_1 -(2\pi\alp)^4
I_{2}^{2}\right]~, \ee \nin where \bea I_1 = \miso g^{\mu\lambda}
g^{\nu\rho} F_{\mu\nu} F_{\lambda \rho}~, \quad I_2 = -\frac{1}{4}
\epsilon^{\mu\nu\lambda\rho} F_{\mu\nu} F_{\lambda\rho}~. \label{i2a}
\eea
Upon replacement in the Lagrangian Eq.~(\ref{bicurved}), one obtains:
\be \mathcal{L}_{\rm DBI} = -\sqrt{{\det}_4 g} \left[e^{-\phi}
  \frac{T_3}{g_{{\rm s}0}} + \frac{1}{2} (2\pi\alp)^2
  e^{-\phi}\frac{T_3}{g_{{\rm s}0}}\, I_1 -\frac{1}{2} (2\pi\alp)^4
  e^{-\phi}\frac{T_3}{g_{{\rm s}0}}\, I_{2}^{2} + \dots +
  \mathcal{L}_{\rm G} \right]~,
\label{expand}
\ee where the $\dots$ indicate higher derivative terms, and
$\mathcal{L}_{\rm G} $ indicates closed-string sector gravity
contributions.  Since gravity is allowed to propagate in the bulk, the
four-dimensional form of the gravitational terms can be obtained from
a projection onto the D3-brane of such terms. In the $\sigma$-model
frame the latter are of the form:
\be\label{gravity} \mathcal{L}_{\rm G} = \sqrt{{\det}_4 g} \,
\frac{V^{(6)}}{g_{{\rm s}0}^2 M_{\rm s}^2} e^{-2\phi}
\left(R(g)-\Lambda_2 + 4 \partial_\mu \phi \partial^\mu \phi + \dots
\right)~, \ee
where $V^{(6)}$ is a volume element of extra (compactified) dimensions
(in units of $\sqrt{\alpha '}$), $\Lambda_2 $ is an effective
cosmological constant term, which may arise from compactification, and
$\phi$ is a four-dimensional dilaton kinetic term. The signature of
$\Lambda_2$ depends on the details of the theory. We define the
four-dimensional bulk-induced gravitational constant $\kappa_0$ as:
\be\label{bigc} \frac{1}{\kappa_0} = \frac{V^{(6)}}{g_{{\rm s}0}^2
    M_{\rm s}^2}~.  \ee
Notice that in this model there are two contributions to the
dilaton-induced dark energy: one from the open string sector, as a
result of the
\be e^{-\phi} \frac{T_3}{g_{{\rm s}0}} \equiv \Lambda_1 (\phi)~,\ee
terms in Eq.~(\ref{expand}), and another one from
\be \frac{V^{(6)}}{g_{{\rm s}0}^2 M_{\rm s}^2} e^{-2\phi} \Lambda_2
\equiv {\tilde \Lambda} (\phi)~,\ee
terms.  Combining Eqs.~(\ref{expand}), (\ref{gravity}), the overall
dark energy contributions would then come from
\be\label{darkenergy} \Lambda_{\rm total~dark~energy} = \Lambda_1
(\phi) + {\tilde \Lambda} (\phi)~. \ee
Since in our approach, $\Lambda_1$ is manifestly of fixed sign
(positive, for positive tension D3-branes $T_3 > 0$), the overall sign
of the induced four-dimensional dark energy $\Lambda_{\rm
total~dark~energy}$ depends on the relative strength of its
constituents. On noting the different scaling of the two types of
vacuum energy contributions $\Lambda_1, \tilde\Lambda$ with the string
coupling $g_{s0}e^\phi$, we observe that it is possible to have a
positive dark energy on the four-dimensional world, for positive
$\Lambda_1 (\phi)+{\tilde \Lambda} (\phi)$, with the $\Lambda_1$
contribution being sub-dominant for weak string couplings at late
times. This is the assumption we shall make when analysing the
consequences of the Eq.~(\ref{expand}) in cosmic structure (galaxy)
formation.  For other phases of the brane universe, the sign of
${\tilde \Lambda} (\phi)$ may change. In fact, in the microscopic models
of D-foam~\cite{foam,foam2} we are considering here, this sign depends
on the ratio of the densities of nearby-to-far-away bulk D-particle
defects, with respect to the brane universe. Indeed, the attractive
flux forces between brane and bulk defects moving in directions
transverse to the brane world, have specific dependence on the
relative distance, which are such that for nearby (far away) bulk
D-particles (lying at distances shorter (longer) than the string
length from the brane universe) there are negative (positive) energy
contributions to the brane vacuum potential energy. We shall come back
to this important point later on in our article.

The effective actions in Eqs.~(\ref{expand}) and (\ref{gravity}) do
not give the complete action that couples dynamical gravity to the
recoil vector field; some further modifications need to be done to the
Dirac-Born-Infeld part, which we shall now discuss.

\subsection{Space-time curvature correction terms in the Dirac-Born-Infeld
action}

We first notice that, in addition to the closed-string
sector gravitational field, the effective action of D-foam may contain
induced target-space curvature contributions, that couple to the
open-string sector, in particular to the DBI Lagrangian
Eq.~(\ref{bicurved}).  Such a situation is known to characterise
simple examples of Dp-branes ($p$ stands for the dimensionality of the
D-brane), fluctuating in the bulk space~\cite{Cheung}.  In fact, the
authors of Ref.~\cite{Cheung} represented the essential features of
such fluctuating branes by considering toy examples of world-sheet
actions for a Dp-brane with tachyonic-type deformations of the form
\be\label{sigma}
S_{\sigma}=\frac{1}{4\pi\alp} \int_{\Sigma} \partial X^M \bar \partial
X_M +\frac{c}{4\pi\alp} \int_{\partial \Sigma}
\left(X^i-Y^i(X^\mu)\right)^2~;  \ee
where \nin $X^M$ with $M=0,\dots,D-1$, are the string coordinates. In this
notation, the index $\mu$ takes values from $0$ to $p$ and corresponds
to the $p+1$ coordinates that obey Neumann boundary conditions, while
the rest obey Dirichlet boundary conditions; the latter are denoted by
$X^i$. The limit $c\rightarrow \infty $, which corresponds to a
conformally invariant point of the theory, ensures that the end-points
of the strings are confined to move on the hyper-surface defined by
$Y^i$. In this limit, a space-time effective action for the $Y^i$s can
be identified with the partition functional of $c$ and $Y^i$:
\be
S(Y^i)= \lim_{c\rightarrow \infty} Z\left(c,Y^i\right)=\int dX^i
dX^\mu e^{-S_\sigma}\,.  \ee
\nin The authors of Ref.~\cite{Cheung} split the string fields in
the usual way in a zero mode plus perturbations, \be X^M=x^M+\tilde
X^M\,, \ee \nin and Taylor expanded $Y^i(X)$ around $Y^i\equiv
Y^i(x)$. They then integrated out of the partition function the
contribution of $\tilde X$, and, after expanding to orders in
$(x^i-Y^i)K^{i}_{\mu\nu}$, with
$K^{i}_{\mu\nu}=\partial_\mu\partial_\nu Y^i$ being the extrinsic
curvature, they integrated the quadratic and quartic terms and arrived
at
\be\label{z} Z(c,Y)=Z(c) \int dx^\mu \sqrt{\det G}\left[ \Biggr. 1
  - \alp \zeta(2) R(G) + {\cal O}(\alp^2) \right]~,
\ee
where $\zeta(2) =\pi^2/6$ and $G_{\mu\nu}=\eta_{\mu\nu} + \partial_\mu
Y^i\partial_\nu Y^i$.

In the flat-brane limit of Ref.~\cite{Cheung}, the induced metric $G$
is the only gravitational contribution.  One, however, may generalise
the background $G$ to an arbitrary metric, $g$, which should then be a
consistent solution to the classical equations of motion for the
gravitational field, obtained from this modified gravity action.  The
presence of the curvature corrections in the partition function
Eq.~(\ref{z}) of the open string, imply a target-space effective
action of the form
\be \label{S-ct} S= - \frac{T_3}{g_{\rm s}}\int d^4 x \sqrt{-\det g}
\left[1- \alp \frac{\pi^2}{6} R + {\cal O} \left( \alp^2 \right)
  \right]~, \ee
which is just the ordinary Einstein-gravity theory (to lowest order in
derivatives), in the presence of a positive cosmological constant
$T_3/g_{\rm s}$. From this point of view, the gravitational (Planck)
constant for this theory is $\kappa = g_{\rm s}/(\alpha ' T_3)$.
However, the reader should also notice that there are curvature
contributions to the effective action, coming from the closed string
sector, of the form Eq.~(\ref{gravity}). In the case of constant
dilatons (as is the case of the model of Ref.~\cite{Cheung}), this
will lead to additional contributions to the effective gravitational
constant in four-dimensions, coming from the closed string sector,
while in the case of time-dependent dilatons, of interest to us here,
one would obtain field-varying gravitational couplings. We shall come
back to this point in the next section.

For the moment we note that coupling a U$(1)$ gauge field background
to the model, Eq.~(\ref{sigma}), corresponds to adding the following
boundary term to the action $S_\Sigma$:
\be \frac{1}{\sqrt{2\pi\alpha'}}\oint_{\partial \Sigma} A_{\mu} (X)
 \partial_\tau X^\nu = \frac{1}{2\sqrt{2\pi\alpha'}} \int_{\Sigma}
 \varepsilon_{\alpha\beta} F_{\mu\nu} \partial^\alpha X^\mu
 \partial^\beta X^\nu~, \ee
\nin where we have used Stokes's theorem, and the indices
$\alpha,\beta$ denote world-sheet indices. The $\sqrt{2\pi\alpha'}$
comes from dimensional considerations.  For constant field strength
backgrounds this is formally equivalent to an antisymmetric
$B_{\mu\nu}$-tensor background.  In general, this is not the
case. This addition changes the effective action Eq.~(\ref{z}) in the
two following ways.
\newline
Firstly, the determinant of the (induced) metric changes as
\be\label{bi2} \sqrt{\det g}\rightarrow \sqrt{\det (g+2\pi\alp F)}
\equiv \sqrt{\det (h)}\,, \ee
where we consider a general metric background $g$, according to our
previous discussion, and not only the induced metric $G$.  As we have
discussed in the previous sub-section, the determinant given in
Eq.~(\ref{bi2}) is the one that appears in the DBI action in a curved
background, Eq.~(\ref{bicurved}).  In fact, the tensor $h_{\mu\nu}$ is
defined as
\be h_{\mu\nu} \equiv
g_{\mu\nu} + 2 \pi\alp F_{\mu \nu}~,
\label{defh}
\ee
and its inverse is denoted by $h^{\mu\nu}$, with $h^{\mu\nu}
h_{\nu\rho} = \delta^\mu_{\,\,\rho}$.
\nin Secondly, as it is well-known from matching with string-amplitude
calculations~\cite{andreev}, in addition to the curvature terms in
Eq.~(\ref{S-ct}), there are contributions involving the gauge field
and its target-space derivatives (in the general case of non-constant
field-strength backgrounds). In the superstring theories we are
interested in (despite the fact that we are only considering the
bosonic parts of the pertinent target-space effective actions), it is
known that such corrections start from structures with
four-derivatives acting on the field strengths, i.e. of the
form~\cite{wyllard}
\be\label{fourder} S_{\rm DBI} = -\frac{T_3}{g_{s}} \int d^4 x
\sqrt{\det (h)} \left[ 1 - \alp \frac{\pi^2}{6} R + \mathcal{O}_{\rm
curv} ({\alpha '}^2) + \mathcal{O}\left((\partial \partial F, \partial
F \partial F )^2\right)\right]~, \ee
where $\mathcal{O}_{\rm curv}({\alpha '}^2)$ indicates terms involving
curvature squared terms. All such higher-derivative terms will be
ignored in our low-energy considerations.

\section{The Four-dimensional induced effective action}\label{secac1}

Following the above discussion, we therefore use the four-dimensional
effective action (on a D3-brane world)
\bea
\label{action1}
S _{\rm eff~4dim} &=& \int d^4x \left[ -\frac{T_3}{g_{s0}} e^{- \phi}
  \sqrt{-\det \left(g + 2\pi\alp F\right)} \Biggl(1 - \alpha R(g)
  \Biggr)\right.\nonumber \\ &&- \left. \sqrt{-g} \,e^{-2\phi} \,
  \frac{1}{\kappa_0} \, {\tilde \Lambda} + \frac{1}{\kappa_0}
  \sqrt{-g}\,e^{-2\phi}\,R(g) + \mathcal{O}((\partial \phi)^2)
  \right]~,\eea
(where $g =\det (g)$) for the gravitational field $g_{\mu\nu}$, the
dilaton field $\phi$ and the gauge field $A_\mu$, with field strength
$F_{\mu\nu}\equiv \partial_\mu A_\nu-\partial_\nu A_\mu $ on a
D3-brane. Note that $\kappa_0$ is defined in Eq.~(\ref{bigc}), and
$\alpha = \alp \zeta (2) = \alp \pi^2/6$ in the example of
Ref.~\cite{Cheung}, however we wish to keep the discussion general and
so we use from now on the positive constant $\alpha > 0$ as a
parameter of our model; it has dimensions of length squared.  In this
respect, we also assumed a non-constant dilaton, thereby writing the
string coupling as $g_{\rm s}=g_{{\rm s}0}e^{\phi}$, with $g_{{\rm
s}0} < 1 $ a (perturbatively) weak constant string coupling. The terms
${\cal O}((\partial \phi)^2)$ denote kinetic terms of the dilaton,
which shall not play an important role in our analysis in this work,
for reasons that will be clarified below; hence we do not need to
write them down explicitly.

Using Eq.~(\ref{dbi2}), taking into account that for the background
solutions configurations pertinent to the effects of the D-particle
foam only ``electric-field'' type components $F_{0i}$ are non-trivial
(\emph{i.e}., $I_2 = 0$ in Eq.~(\ref{i2a})) and expanding the square
root in Eq.~(\ref{action1}), while keeping up to four-derivative-order
terms, we finally arrive at the following form of the four-dimensional
effective action on the D3-brane world
\bea\label{4dea} S_{\rm eff~4dim} &=&\int d^4 x \,\sqrt{-g}\, \left[
  -\frac{T_3}{g_{s0}}\,e^{-\phi} - e^{-2\phi} \,\frac{1}{\kappa_0}
  \,{\tilde \Lambda} \right. \nonumber \\ &&- \left. \frac{T_3}{4
    g_{s0}} 2\pi\alp\, e^{-\phi} F^{\mu\nu} F_{\mu\nu} + \alpha
  \frac{T_3}{4 g_{s0}} 2\pi\alp\, e^{-\phi} F^{\mu\nu} F_{\mu\nu}
  R(g) \right. \nonumber \\ &&+ \left. \left(\alpha \frac{T_3}{g_{s0}}\,
  e^{-\phi} + \frac{1}{\kappa_0} \, e^{-2\phi} \right)\, R(g) +
  \mathcal{O}\left((\partial \phi)^2\right) \right]\,.  \eea
In view of the mis-match between open and closed string sectors, we do
observe from Eq.~(\ref{4dea}) that the coefficients of the Einstein
scalar curvature term $R$ are not of the usual Brans-Dicke type.

We shall base our subsequent analysis upon Eq.~(\ref{4dea}).  We may
proceed by redefining the space-time metric in such a way that the
bulk-induced Einstein curvature term $(1/\kappa_0) \,
e^{-2\phi}\, R(g)$ acquires a canonically normalised term without
dilaton couplings, that is we redefine: 
\begin{equation}\label{redef}
g_{\mu\nu} \rightarrow {\tilde g}_{\mu\nu} = e^{-2\phi} g_{\mu\nu} ~,
\end{equation}
which defines the so-called Einstein-frame metric~\cite{aben}. In our
conventions, such a redefinition implies for our four-dimensional
brane world:
 \begin{equation}\label{Rtilde}
 {\tilde R}({\tilde g}) = e^{2\phi} \, \left[ R (g) -
   \frac{6}{\sqrt{-g}}\, \partial_\mu \left(\sqrt{-g} \, g^{\mu\nu} \,
   \partial_\nu \phi \right) + 6 \, g^{\mu\nu} \partial_\mu \phi \,
   \partial_\nu \phi \right]~.
 \end{equation}
Upon combining Eq.~(\ref{redef}) with a canonical normalisation of the
kinetic terms of the gauge fields, by absorbing the factor $(2\pi\alp e^{-\phi_0}
T_3/g_{s0})^{1/2}$ into a redefinition of the gauge potential $A_\mu$,
we finally obtain the following effective action in the Einstein frame
$S^E_{\rm eff~4dim}$ (for brevity from now on we omit the tilde
notation from the Einstein metric ${\tilde g}_{\mu\nu}$):
\bea\label{4deafinal} && S^E_{\rm eff~4dim} = \int d^4 x \,\sqrt{-g}\,
\left[ -\frac{T_3}{g_{s0}}\,e^{3\phi} - e^{2\phi} \,
  \frac{1}{\kappa_0} \,{\tilde \Lambda} - \frac{1}{4}\,
  F^{\mu\nu} F_{\mu\nu} \right. \nonumber \\ && +\left. \alpha
  \frac{1}{4} F^{\mu\nu} F_{\mu\nu} e^{-2\phi} R(g) + \left(\alpha
  \frac{T_3}{g_{s0}} \, e^{\phi} + \frac{1}{\kappa_0}\right) \, R(g) -
  \frac{3}{2} \, \alpha \, e^{2\phi} g^{\mu\nu}\, \partial_\mu
  (F_{\alpha\beta}\, F^{\alpha\beta}) (\partial_\nu \phi ) +
  \mathcal{O^\prime}\left((\partial \phi)^2 \right) \right]\, + S_m,
\eea
up to terms proportional to the square of dilaton gradients
$\partial_\mu \phi$, denoted by $ \mathcal{O^\prime}\left((\partial
\phi)^2 \right)$, which we shall ignore, as we have already
mentioned. The reader should have noticed, however, that terms
containing one derivative of the dilaton have been kept, since they
will yield non-trivial contributions to the dilaton equation of motion
even in the case of constant dilaton, we are considering.  In
Eq.~(\ref{4deafinal}) we have added the matter sector action $S_m$
which we do not specify.  Notice that for effectively constant dilaton
$\phi = \phi_0$ we are interested in here, the gravitational constant
(that is the coefficient of the Einstein term) reads
\begin{equation}\label{effkappa}
\frac{1}{\kappa} \equiv \frac{1}{\kappa_0} + \alpha \frac{T_3}{g_{s0}}
\, e^{\phi}~.
\end{equation}
This is the action we shall use from now on in order to analyse the
effects of the vector perturbations in structure formation in this
stringy universe. Notice that, for the sake of brevity, we used above
the same symbol $F$ for the field strength of the canonically
normalised gauge fields.

Before proceeding to discuss the role of perturbations, let us
discuss first the background configurations about which we shall
perturb.  First we may assume that at early epochs of the universe,
which will not be the subject of our study here, there is a de~Sitter
phase of the brane universe, in which the population of D-particle
defects is sufficiently dilute, so that any recoil vector field
strength $F$ contributions to the effective action are
\emph{negligible}. From Eq.~(\ref{4deafinal}) we observe that apart
from the term $-(T_3/g_{{\rm s}0})\,e^{3\phi}$, the rest of the terms
in the effective action have the form of a low-energy gravitational
action coming from the closed sector of string theories, within
graviton, dilaton ($\phi$) and tachyon ($T(x)$) backgrounds, the
latter one corresponding to initial cosmological instabilities. Such
de~Sitter phases have been argued to be consistent with world-sheet
conformal invariance to all orders in $\alpha '$~Ref.~\cite{AKM}. In
fact, the de~Sitter phase corresponds to a relatively brief
cosmological era, during which the amplitudes of the (classical)
tachyon and dilaton fields are proportional to each other, $T(x) +
\phi (x) =0$.  Once such a field alignment is lost, due to a decay in
time of the tachyon instability, the universe exits dynamically the
de~Sitter phase.  It is important to notice that in the solution of
Ref.~\cite{AKM}, the dilaton in the Einstein frame also assumes in
such a phase a logarithmic Robertson-Walker-frame time dependence,
$\phi = \phi_0 {\rm ln} t$, $\phi_0 < 0$. This implies that the
afore-mentioned open-string term $-(T_3/g_{{\rm s}0})\,e^{3\phi}$ is
negligible compared to the rest of the terms in the effective action
Eq.~(\ref{4deafinal}) and thus it can be approximated by the one
derived from closed strings. This justifies the existence of an
approximately de~Sitter phase in our context.

At late eras, for instance those corresponding to red-shifts around $z
\simeq 2$, relevant for structure formation which we are interested
in, the D-particle populations, especially near galactic centres, are
significant, thereby leading to the presence of recoil vector field
$F$-contributions to the effective action. In such eras, the
background configuration for the D-particle recoil operators which
satisfies the (logarithmic) conformal algebra on the world-sheet of
the string~\cite{gravanis}, and is therefore a consistent solution of
the equations of motion of the target-space effective action, involves
approximately constant dilatons and spatially flat Robertson-Walker
metric backgrounds in the Einstein frame.

This, in conjunction with Eq.~(\ref{metric}), gives us that
\be\label{finalsol} \phi = \phi_0 ={\rm const}~, \quad g_{00} =
-1+2\sinh(\phi_0)u_0^2~, \quad g_{ij} =
a^2(t)(\delta_{ij}+2\sinh(\phi_0)u_iu_j)~.
\end{equation}
However, in what follows we shall redefine the coordinates in such a
way so that the standard FLRW metric is used as our space-time
background in order to make contact with observations.

The configuration for the D-particle recoil operator in such
backgrounds reads~\footnote{For completeness we note at this stage
  that, even if we worked with the metric Eq.~(\ref{finalsol}), the
  factors depending on ${\rm sinh}(\phi_0) \, u_i^2$ can be ignored
  below, because they give rise to terms cubic in $u_i$ or
  higher. However, by performing the above-mentioned coordinate
  transformations, we make sure that no extra factors involving the
  dilaton $\phi_0$ would hamper the comparison with the data.}:
\begin{equation}\label{recoilfrw}
V = \int_{\partial \Sigma} g_{ij}(t) y^j(t) \Theta (t - t_{\rm impact})
\partial_n X^i~,
\end{equation}
where $\partial_n$ denotes the normal derivative on the world-sheet
boundary $\partial \Sigma$, $t$ is the cosmic time, $g_{ij}(t) =
a^2(t)$ is the spatial part of the FLRW metric and
\begin{equation}\label{geodesic}
y^i(t) = \frac{v^i} {1 - 2p}\left(t \frac{a^2(t_{\rm impact})}{a^2(t)}
- t_{\rm impact} \right) + \mathcal{O}(t^{1-4p}) \simeq
-\frac{v^i}{1-2p} t_{\rm impact} ~, \quad \ \ \mbox{with}\ \ t \gg
t_{\rm impact}~,
\end{equation}
is the D-particle recoil trajectory, with ${\vec v}$ the D-particle
recoil three velocity in the co-moving frame and $t_{\rm impact}$ the
moment of impact of the matter string with a single D-particle. Let us
note that this expression stems from the geodesic trajectories of a
recoiling D-particle in the metric background~\cite{gravanis}.

In our study we consider the interaction of strings with populations
of D-particles, in which case $t_{\rm impact}$ is an averaged time,
considered in what follows as a free constant.  Let us, for
convenience, absorb the constant $t_{\rm impact}/(1 - 2p) $ in the
definition of velocities, which also absorbs the normalisation factors
$(2\pi\alp e^{-\phi_0} T_3/g_{s0})^{1/2}$ employed above in order to normalise the
Maxwell terms in Eq.~(\ref{4deafinal}).  This is equivalent to
considering a vector gauge field background with spatial components of
the form:
\begin{equation}\label{spacevector}
A_i (t) = -a^2(t) \, \delta_{ij}v^j = - a(t) \delta_{ij} v_{\rm
(phys)}^j ~,
\end{equation}
where $v_{\rm (phys)}^j = a(t) v^j $ , $j=1,2,3$ is the physical
(``local'') recoil velocity.  The above analysis holds for low three
velocities; otherwise, one should replace in Eq.~(\ref{spacevector})
the three velocity by the spatial components of the four-vector of the
velocity, $u^i = \gamma v^i$ with $\gamma = 1/\sqrt{1 - v^i v^j
g_{ij}}$.

The background Eq.~(\ref{spacevector}) corresponds to a field strength
of ``electric field'' type:
\begin{equation}\label{elfield}
F_{0i} (t) =  -2\dot{a}(t)a(t)  u_i~.
\end{equation}
where the over-dot denotes the derivative with respect to cosmic time
$t$.

We remind the reader that in this formalism~\cite{kogan}, conformal
invariance of the stringy $\sigma$-model does not restrict the form of
the temporal component of the four-vector field $A_\mu$, apart from
the fact that homogeneity and isotropy require it to depend only on
the cosmic time, $A_0 (t)$.  This freedom allows us to covariance
the background vector field as follows:
\begin{equation}\label{vectorcov}
A_\mu = - a^2(t) u_\mu = - a(t) \, u^{\rm phys}_\mu ~.
\end{equation}
In the Roberson-Walker background of Eq.~(\ref{finalsol}), the
physical (``local'') four velocity $u^{\rm phys}_\mu \equiv a(t) \,
u_\mu $ obeys the Minkowski-flat constraint:
\bea\label{constr1} u_{\mu}^{\rm (phys)}u_{\nu}^{\rm (phys)} \,
\eta^{\mu\nu} = - |\rm constant | = - \frac{|T_3|}{g_{s0}} 2\pi
\alpha ' e^{-\phi_0}< 0~, \eea
from which it follows that the vector field has the following
time-like constraint in our Robertson-Walker background:
\bea\label{constraint} A_{\mu}A_{\nu}g^{\mu\nu} = -
\frac{|T_3|}{g_{s0}} 2\pi \alpha ' e^{-\phi_0}< 0~.  \eea
Unlike the case of Ref.~\cite{bekenstein}, however, in our model the
vector field has non-trivial space-like components, proportional to
$a^2(t)u_i$.  When we consider perturbations, we shall maintain the
above constraint, Eq.~(\ref{constraint}), which is implemented in a
path integral via an appropriate Lagrange multiplier $\lambda (x)$
term in the effective action, namely
\begin{equation}\label{lagrange}
S_{\rm eff-Lagrange} = \int d^4 x \sqrt{-g} \, \lambda (x) \,
\left(A_\mu A^\mu + \frac{|T_3|}{g_{s0}} 2\pi \alpha ' e^{-\phi_0}\right)~,
\end{equation}
which notably does not couple to the dilaton field. The above should
be added to the right hand side of Eq.~(\ref{4deafinal}).

We next remark that, upon averaging over foam populations, one assumes
the following Gaussian stochastically fluctuating configurations:
\be\label{gauss} \ll u_i \gg = 0 \ \ \ \mbox{and}\ \  \ll u_i u_j
\gg = \sigma_0^2 \delta_{ij}~, \ee
where $\ll \dots \gg$ indicates appropriate averages over D-particles
populations in the foam.
Thus, we observe that only upon averaging
over the isotropic foam, we obtain vanishing of the spatial
components of the $A_\mu$ field and of the ``electric field'',
Eq.~(\ref{elfield}), namely
\begin{equation}\label{averages}
\ll A_i \gg =0\ \ \ \mbox{and} \ \  \ll F_{0i} \gg =0~.
\end{equation}
Nevertheless, in our case caution should be exercised to take these
averages only at the very end of the computations, since even powers
of $u_i$ yield non-zero population averages. The parameter
$\sigma_0^2$ is free in our model, albeit small, to be constrained by
the data. In general, $\sigma_0^2$ is cosmic time dependent since it
is a function of the (bulk space) density of the foam, which may vary
for different cosmological eras~\cite{foam,foam2}.  In general,
assuming a no-force condition among the D-particles~\cite{foam2},
which characterises D-foam models derived from super-membrane
theories~\cite{foam}, implies that the density of foam scales as
``dust'' with the cosmological scale factor, $a^3(t)$, so that to a
good approximation over the time period of structure formation we are
interested in here, we may take
\begin{equation}\label{sigma_0dust}
\sigma_0^2 (t) = \frac{\beta}{a^3(t)}  
\frac{2\pi\alpha' |T_3|e^{-\phi_0}}{g_{s0}}~, \quad \beta = {\rm constant}
\, > \, 0~.
\end{equation}
Here the factor $ 2\pi\alpha' e^{-\phi_0}|T_3|/g_{s0}$ comes from
the constants which were previously absorbed into the definitions
of the recoil velocities.

Some important remarks are due at this juncture. The above
considerations are in agreement with the isotropy and homogeneity of
the observed Universe. However, for completeness we should mention
that when considering the effects of our model at galactic scales, it
is possible that certain inhomogeneities in the density of the
D-particles may occur in such a way that D-particle populations
dominate the haloes of galaxies, thus playing a role analogous to
dark matter.  Indeed, due to their localised nature (as contrasted
with the space filling D-brane universes ), D-particles may also be
considered as massive excitations of the vacuum with mass of order
$M_{\rm s}/g_{\rm s}$. Since there is a no-force condition among them,
such particles are ideal to play the role of super-weak dark matter
(termed \emph{D-matter}~\cite{shiu}), and if they are sufficiently
light, as is the case of low string scales $M_{\rm s}$ of order TeV,
and $g_{\rm s} = \mathcal{O}(1)$, they may have implications at
colliders, where they can be produced by the collision of Standard
Model particles~\cite{shiu,foam2}.  However, as discussed in
Ref.~\cite{foam2}, and reviewed briefly above, in our model of D-brane
universes propagating in bulk spaces punctured by D-particle
populations, there are contributions to the vacuum energy on our brane
world by the bulk D-particles, of mixed signature, that depend on the
relative distance of the D-particle from the brane world. Negative
(positive) energy contributions are due to bulk D-particles that lie
at distances smaller (larger) than the string scale from the brane. As
a consequence of such contributions, the density of D-particle
populations that are trapped on our brane world is not restricted by
the requirement of avoiding over-closure of the Universe and can thus
be an arbitrary phenomenological parameter. In this sense, scenarios
in which various mass density profiles for D-matter can exist in the
haloes of galaxies, playing a role analogous to dark matter, are
compatible with current cosmology. When we adapt our model above,
Eq.~(\ref{4deafinal}), to such a situation, we observe that, in view
of our coupling of the vector recoil field to the curvature of
space-time, averaging over populations of D-particles in the galactic
haloes, will result, at a background level, in extra contributions of
the recoil term $\ll e^{-3\phi} \, F_{\mu\nu} \, F^{\mu\nu} \gg$ to an
``effective'' gravitational ``constant'' , proportional to $\sigma_0$,
which in inhomogeneous scenarios may depend on the particular galaxy:
\begin{equation}
\frac{1}{\kappa_{\rm eff}} \equiv \frac{1}{8\, \pi \, G_{\rm eff}} =
\frac{1}{\kappa} + \frac{\alpha}{4} \ll e^{-3\phi} \, F_{\mu\nu} \,
F^{\mu\nu} \gg~,
\end{equation}
where $\kappa$ is defined in Eq.~(\ref{effkappa}) and plays the role
of the four-dimensional Newton's constant $G_{\rm N}$ in our stringy
framework. Such an effective gravitational constant will lead to
Einstein equations coupling the matter and gravitational systems with
a space and time varying effective $G_{\rm eff}$ which can then affect
cosmological considerations. Although such scenarios are worth
pursuing further, especially as providing alternatives to conventional
dark matter (in view of the absence of any concrete particle physics
evidence for dark matter at present), nevertheless in the current
discussion we shall restrict ourselves to homogeneous and isotropic
densities of D-particles on our brane universe, which satisfy
Eq.~(\ref{sigma_0dust}). Our aim in this paper is to investigate the
effect of the recoil vector field to the growth of galaxies and not to
provide realistic alternatives to dark matter scenarios. Our
super-membrane effective field theory has its own dark matter,
provided, for instance, by the supersymmetric partners of the
low-energy effective action. Nevertheless, we hope to come back
to such interesting variants of our model in a future work.

In the next section we proceed to consider perturbations around the
background solution, Eq.~(\ref{finalsol}), and discuss their relevance
to structure growth. To compare our analysis with that of
Ref.~\cite{dodelson}, on the importance of the vector field for
structure growth, we neglect in what follows fluctuations of the
dilaton field, assuming them to be suppressed as compared to the
gravitational and vector perturbations. This is consistent with the
phenomenology of low-energy physics at late eras of the universe,
given that such dilaton fluctuations would affect the values of the
string coupling $e^\phi$, and through it of all the coupling constants
of the string-inspired effective theory, which would not be acceptable
from a particle physics viewpoint.

\section{Background Configuration and Equations of Motion}
\label{bkcgd}
 By varying the effective action given in Eq.~(\ref{4deafinal}) with
 respect to the vector, metric and dilaton fields we obtain
 respectively~\footnote{We set to zero the terms involving derivatives
 of the dilaton only at the end of the variations, since the latter is
 assumed constant $\phi_0$ and absorbing the \emph{constant} factors
 $e^{-\phi_0/2}$ into a normalisation of the gauge potential $A_\mu$,
 so that in this case the kinetic term of the recoil gauge field
 assumes a canonically normalised Maxwell form. The constraint
 Eq.~(\ref{constraint}), is then modified appropriately, however the
 dilaton equation remains unaffected, since the dilaton does not
 couple to the constraint term, as already mentioned.}:
\be\label{veom} \left[F_{\nu\mu}\left(1-\alpha
  e^{-2\phi_0}R\right)\right]^{;\nu}+2 \lambda (x) A_\mu \ = 0~, \ee
  \bea\label{geom} && \left[\frac{1}{4}\,
  F^{\alpha\beta}F_{\alpha\beta} + \frac{{\tilde \Lambda}\,
  e^{2\phi_0}}{\kappa_0} + \frac{T_3e^{3\phi_0}}{g_{s0}} - \lambda (x)
  (A_\alpha \, A^\alpha + \frac{|T_3|}{g_{s0}}2\pi \alpha ' \,
  e^{-\phi_0}) \right]\frac{g_{\mu\nu}}{2} \nn &&
  +\left(R_{\mu\nu}-\frac{1}{2}g_{\mu\nu}R\right)\left[\frac{1}{\kappa_0}
  + \alpha \, \frac{T_3}{g_{s0}} \, e^{\phi_0} + \frac{\alpha
  e^{-2\phi_0}}{4}F^{\alpha\beta}F_{\alpha\beta}\right]
  -\frac{1}{2}g^{\sigma\lambda}F_{\mu\lambda}F_{\nu\sigma}(1-\alpha
  e^{-2\phi_0}R) \nn &&+ \frac{\alpha
  e^{-2\phi_0}}{4}\Biggl\{g_{\mu\nu}\nabla^2\left[F^{\alpha\beta}F_{\alpha\beta}
  \right]-\nabla_\mu\nabla_\nu\left[F^{\alpha\beta}F_{\alpha\beta}\right]\Biggr\}
  = T^{\rm m}_{\mu\nu} - \lambda (x) A_\mu \, A_\nu ~, \eea
and
\be\label{dileq} -\frac{3}{2} \, \alpha \, e^{2\phi_0} \nabla^2
(F_{\alpha\beta} \, F^{\alpha \beta} ) +3\frac{T_3}{g_{s0}}e^{3\phi_0}
+ 2 \frac{{\tilde \Lambda}}{\kappa_0} e^{2\phi_0} - \alpha
\left(\frac{T_3}{g_{s0}}\, e^{\phi_0} - \frac{1}{2}
F_{\mu\nu}F^{\mu\nu} \, e^{-2\phi_0} \right) \, R(g) = 0~, \ee
where $\kappa_0$ is defined in Eq.~(\ref{bigc}), $T_{\mu\nu}^{\rm m} $
is the matter stress tensor and we recall that $\alpha = \alp \zeta
(2) = \alp \pi^2/6 > 0 $ in the D-brane example of
Ref.~\cite{Cheung}. Above, we have assumed for simplicity that the
dilaton coupling to the matter action is not dominant (or equivalently
that the dilaton couples only through exponential couplings of the
form $\exp(\gamma_i \, \phi)$, with $\gamma_i $ constants, to the
various matter fields, such that upon considering the corresponding
dilaton variations, one obtains terms proportional to the Lagrangians
of the various matter-species $i$ that vanish on shell). As our
primary purpose here is to investigate the role of the vector
perturbations on the structure formation and growth, we believe that
such an assumption is reasonable.

Before actually solving the system of equations (\ref{veom}),
(\ref{geom}) and (\ref{dileq}) for the galactic growth and structure
formation epoch of the universe, we consider it as instructive to
discuss the emergence of a de~Sitter phase in this system, which as we
shall argue may indeed characterise \emph{late eras} in the evolution
of this stringy universe, thereby making our model consistent with the
current cosmological observations indicating the dawn of an
accelerating de~Sitter evolution era once again in the history of our
observable Universe.

\subsection{Condensates of the recoil vector field and  late-epoch phases  
of the D-foam universe} 

To demonstrate the existence of a de~Sitter era we first remark that
the dynamics of the system of the recoil vector field is described by
a Dirac-Born-Infeld Lagrangian given in Eq.~(\ref{bicurved}). For such
lagrangians, the higher order interaction terms among the vector field
strength may lead to condensates, in analogy with the gluon
condensates of Quantum Chromodynamics. In fact, such an assumption has
been made in Ref.~\cite{odintsov}, for generic Abelian flux fields
that characterise D-brane excitations, with Dirac-Born-Infeld world
volume actions. The analysis in that work indicated then that under
certain plausible assumptions on the dominance of the quantum effects
(over classical, thermodynamical) on the formation of the condensates,
it is possible to obtain an equation of state for the vector
Dirac-Born-Infeld system that resembles that of a de~Sitter phase,
i.e. a cosmological constant type with $w \simeq -1$.

Let us first review briefly, for instructive purposes, those
results. Consider a generic Dirac-Born-Infeld field, with Lagrangian
Eq.~(\ref{bi}) on a D3-brane world volume. In general, there are two
contributions to the vector field strength $F_{\mu\nu}$ condensates
and one may come from purely quantum vacuum effects,
\begin{equation}\label{vcond}
\langle F_{\mu\nu} \, F^{\mu\nu} \rangle_{{\rm vac} } = {\tilde
  \alpha} (t) ~, \quad \langle F_{\mu\nu} \, F^{\star \, \mu\nu}
\rangle_{{\rm vac} } = {\tilde \beta} (t)~,
\end{equation}
where $F^\star$ indicates the dual tensor, and the condensates may in
general depend on time, but are constant on spatial hyper-surfaces.
The isometry structure of the spatial hyper-surfaces lead the authors
of Ref.~\cite{odintsov} to further assume the following, which we also
adopt here:
\begin{equation}\label{isometry}
\langle F_{0 \, \nu} \, F_{0}^{\,\, \nu} \rangle_{{\rm vac} } =
\frac{\alpha_{\rm t} (t)}{4} \, g_{00}~, \quad \langle F_{i \, 0} \,
F^{0}_{\,\, j} \rangle_{{\rm vac} } = \frac{\alpha_{\rm s} (t)}{4} \,
g_{ij}~,
\end{equation} 
where $i,j$ spatial indices on the three-dimensional volume of the
D3-brane and we have the relations $\alpha_ {\rm t} + 3 \alpha_{\rm s}
= 4{\tilde \alpha} $.  In addition to the quantum effects, one also
has classical thermodynamical effects on the energy density and
pressure of the Dirac-Born-Infeld fluid, which are obtained by averaging
over the spatial volume. If one decomposes the field strength into
``electric-field'', $F_{0i} \equiv E_i$, and ``magnetic-field'' $B_i =
\frac{1}{2} \epsilon_{ijk}\, F^{jk}$ components, one may specify these
classical effects contributions on the condensates~\cite{odintsov}:
\begin{equation}\label{classical}
\langle F_{0 \, \nu} \, F_{0}^{\,\, \nu} \rangle = \langle
\sum_{i=1}^3 \, E_i^2 \rangle ~, \quad \langle F_{i \, 0} \,
F^{0}_{\,\, j} \rangle = - \langle E_i \, E_j \rangle + 2 \langle B_i
\, B_j \rangle~.
\end{equation}
On making the further (natural) assumption~\cite{odintsov} $\langle
E_i \, E_j \rangle = \langle B_i \, B_j \rangle = \mathcal{C} \,
g_{ij}/3$, we observe that the total contribution of both classical
and quantum effects to the vacuum condensates can then be expressed by
the relations:
\begin{equation}\label{totalcond}
\alpha_{\rm t} = {\tilde \alpha} - 4 \mathcal{C}~, \quad \alpha_{\rm
  s} = {\tilde \alpha} + \frac{4}{3}\, \mathcal{C}~.
\end{equation}
Computing the stress tensor of the Dirac-Born-Infeld fluid, Eq.~(\ref{bi}),
for the case of the total condensates one arrives at the following
expressions for energy density $\rho^{\rm DBI}$ and pressure $p^{\rm
  DBI}$~\cite{odintsov}:
\begin{equation}\label{pressure}
\rho^{\rm DBI} = \frac{\lambda}{2} \left(\frac{1 + \frac{{\tilde
      \alpha}}{2} - \frac{\alpha_{\rm t}}{4}}{\sqrt{1 + \frac{{\tilde
        \alpha}}{2} - \frac{{\tilde \beta}^2}{16}}}\right)~, \quad
p^{\rm DBI} = - \frac{\lambda}{2} \left(\frac{1 + \frac{{\tilde
      \alpha}}{2} - \frac{\alpha_{\rm s}}{4}}{\sqrt{1 + \frac{{\tilde
        \alpha}}{2} - \frac{{\tilde \beta}^2}{16}}}\right)~, \quad
\lambda \equiv \frac{T_3}{g_{\rm s}}~.
\end{equation}
Notice that in our case (Eq.~(\ref{bi})) we did not subtract the
determinant of the metric $\sqrt{-g}$ in our definition of the action,
as this plays the role of a cosmological constant contribution,
which we keep. This explains the difference between the
right-hand-side of the above expressions and the corresponding ones in
Ref.~\cite{odintsov}, by constant terms of the form $ \lambda/2$ and
$-\lambda/2$ for $\rho^{\rm DBI}$ and $p^{\rm DBI}$, respectively.

The quantum condensate $\tilde \alpha$ and $\beta$ have not been
specified. The only restrictions come from the positivity of the
corresponding quantities inside the square root in the Dirac-Born-Infeld
action, \emph{e.g}. in the denominators of Eq.~(\ref{pressure}), which
imply the following condition between the various quantum condensates:
\begin{equation}\label{abcond}
\frac{{\tilde \alpha}}{2} \, > -1 + \frac{{\tilde \beta}^2}{16} ~.
\end{equation}
From Eq.~(\ref{totalcond}), it becomes immediately apparent that on
the one hand, if quantum effects are the dominant ones, with ${\tilde
  \alpha} \gg \mathcal{C}$, then $\alpha_{\rm t} \simeq \alpha_{\rm s}
\simeq {\tilde \alpha}$, and hence from Eq.~(\ref{pressure}) we obtain
an equation of state of cosmological constant (de~Sitter vacuum) type,
namely $w^{\rm DBI} \simeq -1 $ independently of the exact form of the
quantum condensate (assuming of course it exists, which is a question
that probably cannot be addressed in a generic manner, as it requires
specific properties of the brane action).  On the other hand, as
demonstrated in Ref.~\cite{odintsov}, when classical effects dominate,
with $\mathcal{C} \gg {\tilde \alpha} $, then Eq.~(\ref{totalcond})
implies  $\alpha_t =-3\alpha_s \simeq -4 \mathcal{C}$ and thus
Eq.~(\ref{pressure}) leads to an ordinary relativistic fluid with
positive energy density and pressure, $\rho^{\rm DBI} \simeq
(\lambda/2) \mathcal{C} \, > 0 $ and $p^{\rm DBI} = (1/3)
\rho^{\rm DBI} \simeq (\lambda/2) \mathcal{C} \, > 0 $.

Hence, the effects of the classical (thermodynamical) contributions to
the condensates are to cause the equation of state of the Dirac-Born-Infeld
fluid to deviate from the value $w_{\rm DBI} \simeq -1 $ by an amount
that depends on the relative strength of those effect with respect to
the quantum ones.

In the context of our study here, the Dirac-Born-Infeld vector field has a
microscopic origin, as describing the dynamics of the recoil degrees
of freedom of the D-particles under their interaction with
electrically neutral matter. In contrast to the case of
Ref.~\cite{odintsov}, here we have the additional couplings of the
space-time curvature to the Dirac-Born-Infeld action, Eq.~(\ref{bicurved}),
whose effects have been ignored in Ref.~\cite{odintsov}. Nevertheless,
as we shall argue, for sufficiently small values of the condensate,
one can still argue on a late de~Sitter era, provided again the
quantum vacuum effects dominate the formation of the condensate.

At this point we would like to emphasise two important remarks.
Firstly, in our case the \emph{classical} effects on the condensates
are just the stochastic effects obtained upon averaging over
statistical populations of D-particles, Eqs.~(\ref{gauss}),
(\ref{averages}) and (\ref{sigma_0dust}). However, the \emph{quantum}
effects have a genuine quantum nature due to the quantum-gravitational
fluctuations of the D-particle defects in the absence of any
matter. The latter are also described by the stretching of virtual
strings between the D-particle and the brane world but they are
non-perturbative, and can only be computed in the strongly coupled
regime of the target space action of the D-particle, therefore only
within the context of M-theory and this is not yet known.
Nevertheless, such effects do exist and in what follows we shall adopt
a phenomenological approach and simply assume (as in the case of
Ref.~\cite{odintsov}) that they can condense, when the D-particle
populations are sufficiently dense on the brane world, and can lead to
the quantum Dirac-Born-Infeld type condensates, Eqs.~(\ref{vcond}),
(\ref{isometry}).

Secondly, in contrast to the generic arguments of
Ref.~\cite{odintsov}, in the case we are studying here, there is the
constraint Eq.~(\ref{constraint}), implemented in an effective action
treatment through the Lagrange multiplier term,
Eq.~(\ref{lagrange}). There are two distinct phases in the problem, as
a result of the quantum treatment of the constraint; we discuss them
below.

\textbf{(a)} The phase in which the constraint is irrelevant,
i.e. $\langle \lambda (x) \rangle = 0$, in which case the vector
Dirac-Born-Infeld field is massless.  This is the case for an era in which
sufficiently dense populations of D-particles cross our brane universe
and they condense due to quantum effects, which are dominant over the
stochastic population effects, Eq.~(\ref{gauss}). In this case, the
quantum effects on the recoil vector field condensate dominate and
one may write, according to our previous discussion and taking into
account the isometry structure Eq.~(\ref{isometry}), that
\begin{equation}\label{cond-v}
\langle F_{\mu \alpha} \, F_{\nu \beta} g^{\alpha \beta}\rangle_{{\rm
    Vac}} = \frac{{\tilde \alpha}(t)}{4} \, g_{\mu\nu}~,
\end{equation}
where the quantum condensate ${\tilde \alpha (t)}$ can be even
negative as we have seen previously (\emph{cf.} Eq.~(\ref{abcond})).
In this phase, the recoil vector field represents quantum fluctuations
of the D-particles in the absence of any matter string
excitations. The nomenclature ``recoil'' may therefore be misleading
but we keep it for uniformity purposes.

\textbf{(b)} The phase in which $\langle \lambda (x) \rangle \equiv
\mathcal{M}^2/2 > 0 $, in which case the vector field becomes
massive. This is the phase where contributions to the growth of
galaxies will come from, and is assumed to characterise the era of
structure formation. Quantum effects on possible condensates are
assumed sub-dominant. One has the \emph{classical} effects on the
condensates Eqs.~(\ref{elfield}), (\ref{gauss}), (\ref{averages}) and
(\ref{sigma_0dust}) that dominate in this case.

\subsubsection{de~Sitter phase and quantum vacuum condensates} 
We concentrate on the phase \textbf{(a)}, which as we shall
demonstrate, is compatible with a late-era de~Sitter phase for our
Universe, even after including the coupling of the target space-time
curvature to the Dirac-Born-Infeld terms< Eq.~(\ref{action1}), which was
ignored in the analysis of Ref.~\cite{odintsov}.  To this end, we
concentrate on the case where the condensate ${\tilde \alpha} $ is
\emph{constant} in time, \emph{small } in magnitude, while the
condensate ${\tilde \beta} = 0$, so that an expansion up to order
${\tilde \alpha} $ in the effective action is sufficient.  To this
order our perturbative equations (\ref{veom}), (\ref{geom}) and
(\ref{dileq}) suffice.  Consider the dilaton equation (\ref{dileq})
first. Upon assuming the formation of quantum condensates,
Eq.~(\ref{cond-v}), we observe that this equation implies that the
scalar curvature of the space-time reads
\begin{equation}\label{desitter}
R = \frac{1}{\alpha \, \left(\frac{2\, T_3}{g_{{\rm s}0}} \, e^{3\phi_0} -
  {\tilde \alpha}\right)} \, \left[\frac{6}{g_{{\rm s}0}} \, T_3 \,
e^{5\phi_0} + \frac{4 {\tilde \Lambda}}{\kappa_0} \, e^{4\phi_0}
\right]~.
\end{equation}
This can be a \emph{positive constant}, as appropriate for a de~Sitter
space-time, provided
\begin{equation}\label{condcond}
\frac{2\, T_3}{g_{{\rm s}0}} \, e^{3\phi_0} - {\tilde \alpha}\, > \, 0 ~.
\end{equation}
If we assume a de~Sitter maximally symmetric form for the Riemann
curvature tensor, corresponding to a Hubble constant, $H_{\rm I}={\rm
  const} > 0$, namely,
$$R_{\mu\nu\rho\sigma} = H_{\rm I}^2 \left( g_{\mu\rho} \, g_{\nu\sigma} -
g_{\mu\sigma} \, g_{\nu\rho} \right)~, $$ we obtain from
Eq.~(\ref{desitter}) the following relation:
\begin{equation}\label{hi1}
H_{\rm I}^2 = \frac{R}{12} = \frac{1}{\alpha \, \left(\frac{2\,
    T_3}{g_{{\rm s}0}} \, e^{3\phi_0} - {\tilde \alpha}\right) } \,
\left[\frac{T_3 }{2g_{{\rm s}0}} \, e^{5\phi_0} + \frac{{\tilde
    \Lambda}}{3\kappa_0} \, e^{4\phi_0} \right] \, > 0~.
\end{equation}
From the graviton equation (\ref{geom}), ignoring the matter
contributions $T^m_{\mu\nu}$, using Eq.~(\ref{cond-v}), and
considering the phase $\langle \lambda \rangle =0$ we obtain
\begin{equation}\label{hi2}
H_{\rm I}^2 = \frac{1}{6} \, \left[\frac{\frac{T_3 }{g_{{\rm s}0}} \, e^{3\phi_0}
  + \frac{{\tilde \Lambda}}{\kappa_0} \, e^{2\phi_0}
}{\frac{1}{\kappa_0} + \alpha \, \left(\frac{T_3}{g_{{\rm s}0}}\, e^{\phi_0}
  - \frac{{\tilde \alpha}}{4}\, e^{-2\phi_0} \right)} \right]~.
\end{equation}
From Eqs.~(\ref{hi1}) and (\ref{hi2}) we finally obtain the value of
the condensate that guarantees a de~Sitter geometry in this phase, namely
\begin{equation}\label{condvalue}
{\tilde \alpha} = \frac{4\, e^{2\phi_0}}{\alpha \, \kappa_0} \,\,
\left[\frac{ 1 + \alpha \, \kappa_0 \, \frac{T_3}{g_{{\rm s}0}}\,
    e^{\phi_0} \, \left(1 - \frac{1}{3} \mathcal{G} \right)}{1 -
    \frac{2}{3} \, \mathcal{G}}\right] ~, \qquad \mathcal{G} \equiv
\frac{\frac{ T_3 }{g_{{\rm s}0}} \, e^{\phi_0} + \frac{{\tilde
      \Lambda}}{\kappa_0}}{\frac{T_3 }{2\, g_{{\rm s}0}} \, e^{\phi_0}
  + \frac{{\tilde \Lambda}}{3\, \kappa_0}} ~.
\end{equation}
Thus, the condensate is proportional to the
factor 
\be \frac{e^{2\phi_0}}{\alpha \, \kappa_0 } =
\frac{6V^{(6)}}{\pi^2(g_{{\rm s}0}e^{-\phi_0})^2}~, \ee 
using Eq.~(\ref{bigc}) and adopting the value of $\alpha$ that appears
in the example of Ref.~\cite{Cheung}.  This factor depends on the size
of the compactification volume.

For mathematical consistency of our solution, the condensate ${\tilde
  \alpha}$ and the curvature, \emph{i.e}. the Hubble constant $H_{\rm
  I}$ (\ref{hi1}) have to be sufficiently small, which is easily
guaranteed from Eq.~(\ref{condvalue}) for sufficiently large negative
values of $\phi_0$.  Such large negative values of the dilaton may
characterise late epochs of the universe. For instance, in the linear
(run-away) dilaton scenario, in terms of the FLRW time in the Einstein
frame, the dilaton assumes large negative values, $\phi \sim -{\rm
  ln}\, t \gg -1 $ for large times $t \to \infty$, and its variation
is suppressed by ${\dot \phi} \sim - 1/t \to 0$, so it may be
considered approximately constant. This could be a phase where the
above conditions for the formation of small condensates are satisfied,
in which case the universe will enter a late de~Sitter phase.

From our microscopic D-particle foam point of view, such a late phase
may occur when matter in the brane is sufficiently dilute due to the
brane world cosmic expansion, and the brane universe passes again a
bulk area with sufficiently dense D-particle populations. The
classical recoil effects of the D-particles, due to their interaction
with matter strings on the brane, become then sub-dominant to their
\emph{quantum vacuum fluctuation} effects. In fact, as already
mentioned previously, in our microscopic models of
D-foam~Refs.~\cite{foam,foam2}, in such densely populated bulk areas,
the bulk cosmological constant ${\tilde \Lambda} $ may be negative,
\begin{equation}\label{dense}
{\tilde \Lambda} = - |{\tilde \Lambda} | \, < \, 0 \quad {\rm
  for~dense~populations~of~nearby~bulk~D-particles}~,
\end{equation}
since the negative contributions to the brane vacuum energy from the
nearby bulk D-particles dominate. It is not inconceivable then,
although admittedly this requires some fine tuning, that the bulk
density populations are such that in order of magnitude
 \begin{equation}
 \mathcal{G} \gg -1 \qquad, \quad {\rm e.g} \quad \frac{T_3}{g_{s0}}\,
 e^\phi_0 \sim \frac{2\, |{\tilde \Lambda} |}{3\, \kappa_0}~,
 \end{equation}
in which case Eq.~(\ref{condvalue}) implies:
\begin{equation}\label{condvalue2}
{\tilde \alpha} \simeq 2\, e^{2\phi_0} \, \frac{T_3}{g_{s0}}\,
e^{\phi_0} \sim e^{2\phi_0 }\, \frac{4\,|{\tilde \Lambda} |}{3\,
  \kappa_0}~,
\end{equation}
and the condensate ${\tilde \alpha} \ll 1$ (as required for
consistency of the approach) for $|{\tilde \Lambda} |/\kappa_0 \ll 1$
and \emph{finite} values of $\phi_0$.

In such a case, the formation of the condensate may be understood as
\emph{stabilising} the brane vacuum. Indeed, as follows from
Eq.~(\ref{geom}), upon the formation of constant \emph{quantum vacuum}
condensates ${\tilde \alpha}$, the dark energy terms in the brane
effective action assume the form
\begin{equation}
{\rm Brane~Dark~Energy} \, \sim \, {\tilde \alpha} + e^{2\phi_0}
\left(\frac{T_3}{g_{s0}}\, e^{\phi_0} - \frac{|{\tilde \Lambda}
  |}{\kappa_0} \right) ~.
\label{de}
\end{equation}
 In the absence of a condensate, the brane vacuum energy would be
 \emph{negative}, of order $-|{\tilde \Lambda} |/\kappa_0$ and his
 would indicate an instability of the vacuum. The true stable vacuum
 of the theory would then be the one in which the condensate forms,
 with the value Eq.~(\ref{condvalue2}), which implies a
 \emph{positive} (de~Sitter-type) \emph{vacuum energy},
 Eq.~(\ref{de}), of order $e^{2\phi_0}\, \frac{|{\tilde \Lambda}
 |}{\kappa_0 } \, > \, 0$, for any value of $\phi_0$. This would
 guarantee the validity of our arguments on a D-vacuum-induced
 de~Sitter phase even in the case of phenomenologically relevant
 values of the string coupling, reproducing the standard model
 couplings today.  However, in the context of our microscopic model
 derived from strings, this case would correspond
 (\emph{cf}. Eq.~(\ref{desitter})) to a large Hubble parameter (and
 hence curvature) of order $M_{\rm s}^2 =1/\alpha '$, so our
 lowest-order approximations would not be valid (this of course does
 not preclude our discussion on the stabilising properties of the
 condensate from being correct in a non-perturbative string
 context). Hence, it seems that, at least within our string effective
 context, one cannot avoid making the assumption of large negative
 values of $\phi_0$ in the discussion of late-era D-foam vacuum
 fluctuations condensates.

\subsubsection{Matter-dominated phase and statistical contributions 
of D-particle populations to vector-field condensates}
We next proceed to discuss the massive phase of the recoil vector
field, where matter effects dominate, and in this sense the
contributions of the classical effects of recoil velocity
fluctuations, obtained after averaging over D-particle populations, to
condensates of the vector field strengths dominate over quantum
fluctuation effects of D-particles.

Assuming only electric-field type backgrounds, which is the background
considered so far, we do have for such classical averages:
\begin{eqnarray}\label{classicalD}
&& \ll F_{0 \, \nu} \, F_{0}^{\,\, \nu} \gg \, = \, \ll \sum_{i=1}^3
  \, E_i^2 \gg \, \equiv \, 3 {\tilde \sigma}^2 (t) > 0~, \nn && \ll
  F_{i \, 0} \, F^{0}_{\,\, j} \gg \, = \, - \ll E_i \, E_j \gg \equiv
  - g_{ij} \, {\tilde \sigma}^2 (t)~,
\end{eqnarray}
where we assumed isotropic recoil fluctuations for simplicity, and we
denoted generically the recoil fluctuations of the field strength
(averaged over D-particle populations) as ${\tilde \sigma}^2(t)$,
which is a function of the cosmic time. We note that scale factor
dependent contributions are included, as in Eqs.~(\ref{elfield}),
(\ref{sigma_0dust}). The metric $g_{ij}$ denotes the FLRW space-time
background.  Using Eqs.~(\ref{elfield}), (\ref{vectorcov}), we get
\begin{equation}\label{tildesb}
{\tilde \sigma}^2(t) = 4H^2 \, a^2(t) {\sigma}_0^2 
= 4\frac{H^2}{a(t)}\, \beta \ \ , \ \  H = \frac{\dot a}{a}~.
\end{equation}
We next notice that, irrespectively of the specific form of the
condensate, in case these classical (statistical) effects are dominant
over the true quantum ones, Eqs.~(\ref{isometry}),(\ref{classicalD})
imply
\begin{equation}\label{isometryD}
\alpha_{\rm t} (t) = - 12 {\tilde \sigma}^2 (t) \, < \, 0~, \quad
\alpha_{\rm s} (t) = + \frac{1}{3} \alpha_{\rm t} (t) = - 4 {\tilde
\sigma}^2 (t)
\end{equation}
and the corresponding Dirac-Born-Infeld analysis Eq.~(\ref{pressure})
yields \emph{positive energy density} \bea \rho^{\rm DBI-recoil} =
\frac{3\, \lambda}{2} \,{\tilde \sigma}^2 (t) \, > 0~,
 \label{birho}
 \eea but \emph{negative pressure } 
\bea\label{bip} p^{\rm DBI-recoil} = - \frac{\lambda}{2} \,{\tilde
 \sigma}^2 (t) = -\frac{1}{3} \, \rho^{\rm DBI-recoil}~, \eea and an
 equation of state \bea\label{bieq} w_{\rm DBI-recoil} \simeq
 -\frac{1}{3}~.  \eea
This is the limiting case for inducing acceleration in the universe,
which is fine given that we are in the matter-dominated phase, where
the decelerating effects of matter cannot be ignored. Hence, all this
result tells us is that the matter-induced recoil effects of
D-particles are not sufficient to create a cosmological-constant-type
accelerating universe, unlike the quantum fluctuation
effects~\footnote{However in our model there are many other factors
that contribute to the equation of state and to acceleration, bulk
D-particles plus closed string sector quantum effects contributing to
brane tension. Moreover, in our analysis in this work, we have ignored
the presence of recoil-induced ``magnetic-field'' contributions.  We
do notice at this stage that for stability reasons on a D-brane
universe, magnetic-field type backgrounds are also necessary. In our
recoil case this is easily achieved, as discussed in
Ref.~\cite{mavromatosD}, by considering angular momentum type
contributions in the world-sheet deformations (we give them here in
flat space-time for brevity), which are still compatible with the
logarithmic conformal algebras,
\begin{equation}
V^{\rm angular~momentum} \propto \int_{\partial \Sigma} u^i
\epsilon_{ijk} X^j \Theta (t - t_0) \partial_n X^k =\ \int_{\Sigma}
\epsilon^{\alpha\beta} \left( u^i \epsilon_{ijk} \Theta (t - t_0)
\partial_\alpha X^j \partial_\beta X^k + \mathcal{O}(\delta
(t-t_0))\right)~,
\end{equation}
and yield magnetic-field type background in target space $B_i \sim u_i
$, with $u_i$ the recoil velocity of the defect. Such backgrounds do
contribute upon performing statistical averages over D-particle
populations, terms $\ll B_i B_j \gg \, \sim \, \ll E_i E_j \gg \, \sim
\, \delta_{ij} \, {\tilde \sigma}^2 (t) $, thereby making the
situation entirely analogous to the one considered in
Ref.~\cite{odintsov} and described briefly above. The result of these
classical contributions then in this case is a fluid for the vector
field describing the dynamics of D-particle-recoil which has
\emph{positive} pressure and energy and thus, in the case the
constraint Eq.~(\ref{constraint}) is ignored, behaves as a
relativistic matter fluid with $p^{\rm DBI} = (1/3) \rho^{\rm DBI} >
0$.  We shall not consider, though, such magnetic field contributions
in what follows but we felt stating their potentially important role
for completeness.}.

We will use Eqs.~(\ref{isometry}), (\ref{isometryD}) in the equations
of motion Eqs.~(\ref{dileq}) and (\ref{geom}), but keeping the matter
stress energy tensor terms $T_{\mu\nu}^{\rm m}$.

Let us first analyse the graviton equation (\ref{geom}). We have
\bea\label{geom_2} && R_{\mu\nu}-\frac{1}{2}g_{\mu\nu} + g_{\mu\nu} \,
8\pi {\rm G}_{\rm eff} \, \Lambda_{\rm eff} = 8\pi {\rm G}_{\rm eff}
\, T^m_{\mu\nu} + 8\pi {\rm G}_{\rm eff} \, \mathcal{F}_{\mu\nu} \eea
whereby 
\bea\label{defeff} {\rm G}_{\rm eff} &\equiv& \frac{1}{8\pi}
\left(\frac{1}{\kappa_0} + \alpha \, \frac{T_3}{g_{{\rm s}0}} \,
e^{\phi_0} + \frac{\alpha
e^{-2\phi_0}}{4}F^{\alpha\beta}F_{\alpha\beta}\right)^{-1}~, \nn \,
\Lambda_{\rm eff} &\equiv& \frac{1}{8}\,
F^{\alpha\beta}F_{\alpha\beta} + \frac{{\tilde \Lambda}\,
e^{2\phi_0}}{2\kappa_0} + \frac{T_3e^{3\phi_0}}{2g_{{\rm s}0}} -
\frac{\langle \lambda (x) \rangle}{2} \, \left[A_\alpha \, A^\alpha +
\frac{|T_3|}{g_{{\rm s}0}}2\pi \alpha ' \, e^{-\phi_0}\right]~, \nn
\mbox{and}\ \ \mathcal{F}_{\mu\nu} &\equiv&
\frac{1}{2}g^{\sigma\lambda}F_{\mu\lambda}F_{\nu\sigma}(1-\alpha
e^{-2\phi_0}R) - \frac{\alpha
e^{-2\phi_0}}{4}\Biggl\{g_{\mu\nu}\nabla^2
\left[F^{\alpha\beta}F_{\alpha\beta}\right]-\nabla_\mu\nabla_\nu
\left[F^{\alpha\beta}F_{\alpha\beta}\right]\Biggr\} - \langle \lambda
(x) \rangle A_\mu \, A_\nu ~, \nn \mbox{with}\ \ \langle \lambda (x)
\rangle &=& -\frac{g_{s0}e^{\phi_0}}{|T_3|4\pi\alpha'}
A^{\mu}\left[F_{\nu\mu}(1-\alpha e^{-2\phi_0}R)\right]^{;\nu}.  \eea
The form of the lagrange multiplier $\lambda$ is found by contracting
the vector equation of motion, Eq.~({\ref{veom}}), with $A^{\mu}$ and
then applying the constraint Eq.~({\ref{constraint}}).  

Next, we use the dilaton constraint Eq.~(\ref{dileq}) to rewrite the
term ${\tilde \Lambda}\, e^{2\phi_0}/\kappa_0$ in $\Lambda_{\rm
eff}$. For convenience we split $\Lambda_{\rm eff}$ into two parts such
that
\bea\label{splitlameff} \Lambda_{\rm eff} &=& \Lambda^{(0)}_{\rm eff}
+ \Lambda^{(1)}_{\rm eff}~,\nn \mbox{with}\ \ \Lambda^{(0)}_{\rm eff}
&\equiv& -\frac{T_3e^{3\phi_0}}{4g_{{\rm s}0}}\left(1-\alpha
e^{-2\phi_0}R\right) - \frac{\langle \lambda (x) \rangle}{2} \,
\left[A_\alpha \, A^\alpha + \frac{|T_3|}{g_{s0}}2\pi \alpha ' \,
e^{-\phi_0}\right]~,\nn \mbox{and}\ \ \Lambda^{(1)}_{\rm eff} &\equiv&
\frac{F^{\alpha\beta}F_{\alpha\beta}}{8}\left(1-\alpha
e^{-2\phi_0}R\right) + \frac{3\alpha
e^{2\phi_0}}{8}\nabla^2\left[F^{\alpha\beta}F_{\alpha\beta}\right]~.
\eea
This allows the separation of those parts of $\Lambda_{\rm eff}$ which
depend on $\sigma_0^2$ terms from those which are
$\sigma_0^2$-independent. Similarly, ${\rm G}_{\rm eff} $ can also be
split as
\bea\label{splitgeff} \, {\rm G}_{\rm eff} &=& \left[G^{(0)}_{\rm eff}
  + G^{(1)}_{\rm eff}\right]~,\nn \mbox{with}\ \ G^{(0)}_{\rm eff}
  &\equiv&{1\over 8\pi} \ \left[\frac{1}{\kappa_0}+\frac{\alpha
  T_3e^{\phi_0}}{g_{{\rm s}0}}\right]^{-1}~,\nn \mbox{and}\ \
  G^{(1)}_{\rm eff} &\equiv& \frac{\alpha
  e^{-2\phi_0}F^{\alpha\beta}F_{\alpha\beta}}{32\pi}\left[\frac{1}{\kappa_0}
  +\frac{\alpha T_3e^{\phi_0}}{g_{{\rm s}0}}\right]^{-2}~. \eea
The constant $G^{(0)}$ plays the role of the gravitational constant
in this scenario.  The background configuration of the fields has
already been stated in Eq.~(\ref{finalsol}) and
Eq.~(\ref{vectorcov}). This then allows us to find the background
solutions to the equations of motion.

\subsection{The Background Solution}
For the background, the configuration of the gauge and graviton fields
has been explained in the earlier sections and was shown to be
\bea
\label{background}
g_{00} &=& -1~, \nn
g_{ij}  &=& a^2(t)\delta_{ij}~,\nn
A_{0} &=& \left[\frac{2\pi\alpha' |T_3|e^{-\phi_0}}{g_{s0}}+u^2a^2(t)\right]^{1/2}~, \nn
A_{i} &=& -a^2(t)u_i~.
\eea 
Furthermore, assuming that the matter in our system can be treated as an ideal
fluid, it is found that
\bea
\label{Pfluid} 
T^{{\rm m}\mu}_{~~~\nu} = (\rho + P)w^\mu
w_\nu+P\delta^\mu_{~\nu}~, 
\eea 
where $\rho$ and $P$ are the density
and pressure of the fluid and $w_\mu$ is its velocity vector field, normailised
to be timelike. Thus
\bea
T^{{\rm m} 0}_{~~~0} &=& -\rho~,\nn
T^{{\rm m} i}_{~~~j} &=& P\delta^{i}_{~j}~.
\eea
With the background configuration defined, we begin looking at the 
background solutions 
by examining the vector equation Eq.~(\ref{veom}). 
By contracting Eq.~(\ref{veom}) with $A^{\mu}$ is possible to
find the form of the Lagrange multiplier in the background. This is useful
as it can then be used in the other equations of motion when calculating 
their background forms. Thus from Eq.~(\ref{veom}) we can see that
\bea
\lambda(x) \frac{2\pi\alpha' |T_3|e^{-\phi_0}}{g_{s0}} 
= -\frac{A^\mu}{2}\left[F_{\nu\mu}
(1-\alpha e^{-2\phi_0}R)\right]^{;\nu}~. 
\eea
With our background and after taking the vev of the recoil velocity in 
accordance with Eq.~(\ref{gauss}), it is found that the above relation
gives
\bea
\label{lambda}
\langle \lambda \rangle \frac{2\pi\alpha' |T_3|e^{-\phi_0}}{g_{s0}} =
3\sigma_0^2 a^2
\left[\frac{\ddot{a}}{a}+2H^2- 6\alpha
e^{-2\phi_0}\left(\frac{H\dddot{a}}{a}+\frac{\ddot{a}^2}{a^2}
+\frac{4H^2\ddot{a}}{a}\right)\right]~. \eea
Next we can see that the vector equation of motion, Eq.~(\ref{veom}),
 is linear in the gauge field.
Thus, when looking at the spatial component of the equation,
all the terms will be proportional to $u_i$ and by Eq.~(\ref{gauss})
such terms will be zero so the spatial component of the equation
vanishes. The temporal component remains non zero as $A_0$ from 
Eq.~(\ref{background}) is not proportial to $u_i$. Thus we find that
the temporal component of the vector equation in the background 
yeilds
\bea
0=
3\sigma_0^2 a^2  \left[\frac{g_{s0}}{2\pi\alpha' |T_3|e^{-\phi_0}}\right]^{1/2} 
\left[\frac{\ddot{a}}{a}+2H^2- 6\alpha
e^{-2\phi_0}\left(\frac{H\dddot{a}}{a}+\frac{\ddot{a}^2}{a^2}
+\frac{4H^2\ddot{a}}{a}\right)\right]~. \eea

Next we can look at the dilaton equation, Eq.(\ref{dileq}). This gives
\bea
0=\left[12\alpha\sigma_0^2 a^2 e^{-2\phi_0}H^2
+\frac{T_3e^{\phi_0}\alpha}{g_{s0}}\right]
\left(\frac{\dda}{a}+H^2\right)
+12\alpha \sigma_0^2 a^2 e^{2\phi_0}
\left[\frac{\dda^2}{a^2}+\frac{H\ddda}{a}+\frac{3H^2\dda}{a}\right]
\eea
However a more convenient way to use the dilaton equation is to 
use it as a constraint in the graviton equation, as was done in
Eq.(\ref{splitlameff}). Thus when moving on to the graviton equation 
we shall be using the dilaton constrained version of this equation.
When considering the background graviton solution, we can write the modified
Einstein equation as
\bea \label{einst} R_{\mu\nu}-\frac{1}{2}g_{\mu\nu}R = 8\pi
    G^{(0)}_{\rm eff}\left[T^{\rm m}_{\mu\nu}-\Lambda^{(0)}_{\rm
    eff}g_{\mu\nu}\right] + T_{\mu\nu}^{\rm DBI}~.  \eea
Here $T_{\mu\nu}^{\rm DBI}$ represents those terms coming from the
effects of the Dirac-Born-Infeld field; it is defined to be
\bea \label{DBIeinst} T_{\mu\nu}^{\rm DBI} = 8\pi G^{(0)}_{\rm
  eff}\left(\mathcal{F}_{\mu\nu}-\Lambda_{\rm
  eff}^{(1)}g_{\mu\nu}\right)-8\pi G^{(1)}_{\rm
  eff}\left(T_{\mu\nu}^{\rm m}-\Lambda^{(0)}_{\rm
  eff}g_{\mu\nu}\right)~. \eea
Note that for the background solution the $\langle \lambda (x)
 \rangle$ term in the definition of $\Lambda_{\rm eff}^{(0)}$ is zero
 upon application of the constraint equation. In this case we can
 write the modified continuity equation as
\bea\label{conteq}
 \nabla^\mu\left[T_{\mu\nu}^{\rm m}-\Lambda_{\rm eff}^{(0)}
   g_{\mu\nu}+\frac{T_{\mu\nu}^{\rm DBI}}{8\pi G^{(0)}_{\rm eff}}\right] =0~. \eea
For the FLRW background we are considering and denoting by $\rho^{\rm
  tot}$ the total density contributions coming from matter ($\rho^{\rm
  m}$), the DBI part ($\rho^{\rm DBI}$) and any dark fluid
  ($\rho^{\Lambda}$), we find, upon assuming a perfect fluid form of
  stress energy tensor, that
\bea \frac{{\rm d}\rho^{\rm tot}}{{\rm d}a}+\frac{3\rho^{\rm
    tot}}{a}+\frac{3P^{\rm tot}}{a} = 0~, \eea 
just as in the standard gravitation case. In order to investigate the
effects of $\rho^{\rm DBI}$ in the matter dominated phase of structure
formation we note the following. The contribution from
$\rho^{\Lambda}$ can be neglected in this phase. Moreover, for
pressureless matter, the total pressure, $P^{\rm tot}$, has only
Dirac-Born-Infeld contributions (including the constraint term
Eq.~(\ref{constr1})) such that $P^{\rm tot} = P^{\rm DBI}$. The
equation of state of the total fluid can then be used leading to
\bea P^{\rm tot} = -{\rho^{\rm DBI}\over 3}~. \label{dbies} \eea
We thus  get the following first order differential equation
\bea \frac{{\rm d}\rho^{\rm tot}}{{\rm d}a}+\frac{3\rho^{\rm tot}}{a}
= \frac{\rho^{\rm DBI}}{a}~, \eea
with solution
\bea\label{totalener} \rho^{\rm tot} =
\frac{\mathcal{C}}{a^3}+\frac{1}{a^3}\int a^2\rho^{\rm DBI} da~, \eea
where $\mathcal{C}$ is a constant to be determined from boundary
conditions. The contribution $\rho^{\rm DBI}$ is
\bea\label{rbi} \rho^{\rm DBI} &=& \frac{T_{00}^{\rm DBI}}{8\pi
  G^{(0)}}\nn &=& 3\dot{a}^2\sigma_0^2\left[1-6\alpha
  e^{-2\phi_0}\left(H^2+\frac{\ddot{a}}{a}\right)\right] -\langle
  \lambda \rangle\frac{|T_3|2\pi \alpha'\, e^{-\phi_0}}{g_{{\rm
  s}0}}+18\alpha
  e^{2\phi_0}\sigma_0^2\left(\dot{a}\dddot{a}+\ddot{a}^2
  +\frac{3\dot{a}^2\ddot{a}}{a}\right)\nn &&+36\alpha
  e^{-2\phi_0}\frac{\dot{a}^2\ddot{a}}{a}\sigma_0^2+ 6\alpha
  e^{-2\phi_0}\dot{a}^2\sigma_0^2\left[\frac{1}{\kappa_0}+\frac{\alpha
  T_3e^{\phi_0}}{g_{{\rm
  s}0}}\right]^{-1}\left[\rho^m-\frac{T_3e^{3\phi_0}} {4g_{{\rm
  s}0}}\left(1-6\alpha e^{-2\phi_0}\left(H^2+{\ddot{a}\over
  a}\right)\right)\right]~. \eea
For the (late) epoch of structure formation, we observe from
Eq.~(\ref{rbi}) that the terms proportional to $\sigma_0^2 \sim a(t)
^{-3} \beta $ (\emph{cf}. Eq.~(\ref{sigma_0dust})) are suppressed by
at least one inverse power of the scale factor $a(t) \sim t^p$. Hence,
to leading order, for a late era we can write
\begin{equation}\label{rbi2}
\rho^{\rm DBI} \simeq -\langle \lambda \rangle\, \frac{|T_3|2\pi
  \alpha'\, e^{-\phi_0}}{g_{{\rm s}0}} +
  \mathcal{O}\left(\frac{1}{a}\right)~.
\end{equation}
The background form of the Lagrange multiplier was calculated
previously in Eq.(\ref{lambda}). Upon examination again we see that the
Lagrange multiplier terms are suppressed by at least one inverse power
of the scale factor. Thus for late time eras, the DBI density
contribution can be taken to be negligible.

The total energy density $\rho^{\rm total} = \rho^{\rm m} + \rho^{\rm
DBI} $ is determined from Eq.~(\ref{totalener}).  Matter dominance
already required that the DBI terms should be suppressed compared to
the standard $1/a^3$ terms, which we have now explicitly shown is the
case.  Thus, to a good approximation the matter dominated era scale
factor, scaling with the FLRW time, can be taken to be $a_{\rm matter}
(t) \sim t^{2/3}$, as in the standard gravitational case. Lastly
then, we can set the constant of integration, $\mathcal{C}$, to be
equal to the ordinary matter density today, $\rho_{{\rm m}0}$. We find
it convenient to work in units of the critical density today
$\rho_{\rm cr} = 1$, in which $\rho_{{\rm m}0}= \Omega_{\rm m}$ in the
standard notation. Thus, to leading order for the era of structure formation
where the effect of the cosmological constant can be neglected,
\begin{equation}
\label{bckgrndrho}
\rho^{\rm tot} = \frac{\Omega_{\rm m}}{a^3} +
\mathcal{O}\left(\frac{1}{a}\right)~.
\end{equation}
Some important comments are now in order concerning the nature of the
$\Omega_m$. In our model, as already mentioned, there are
contributions to $\Omega_m$ from the D-particles (due to their masses)
as well as the conventional dark matter components (if there are any).
However, due to the competing effects of the nearby and far-away bulk
D-particles (\emph{cf}. Eqs.~(\ref{long-1}), (\ref{pot1})), there may
be significant cancellations in the D-particle contribution to
$\Omega_m$.  As we shall discuss in Section \ref{sec:growth}, to
isolate the effects of D-matter we shall consider the extreme case
where $\Omega_m$ is only baryonic and the D-particles play an
important role in growth mainly due to their interaction with
neutral matter, which provides the seed for the appearance of the
recoil vector field and its associated perturbations.

Finally we can write down the explicit form of the density and pressure
equations coming from the dilaton constrained graviton equation of motion.
Here we shall not assume any particular era of the universe and give the
full general equations, unlike previously when considering the continuity equation
where only the matter domination era was considered and the effect of 
$\rho_{\Lambda}$ could be neglected. Below we have also incorporated
the form of the Lagrange multiplier, Eq.~(\ref{lambda}).
\bea\label{rpi}
\rho &=& 3H^2\left[\frac{1}{\kappa_0}
+\frac{\alpha T_3 e^{\phi_0}}{g_{s0}}\right]+
\frac{T_3e^{3\phi_0}}{4g_{s0}}
\left[1-6\alpha e^{-2\phi_0}\left(H^2
+\frac{\dda}{a}\right)\right]\nn
&&+\sigma_0^2a^2\left[3H^2+\frac{3\dda}{a}
-18\alpha e^{-2\phi_0}\left(\frac{\dda^2}{a^2}+\frac{H\ddda}{a}
+\frac{5H^2\dda}{a}\right)-18\alpha e^{2\phi_0}\left(\frac{\dda^2}{a^2}
+\frac{H\ddda}{a}
+\frac{3H^2\dda}{a}\right)\right]\nn
-P &=& \left(H^2+\frac{2\dda}{a}\right)\left[\frac{1}{\kappa_0}
+\frac{\alpha T_3 e^{\phi_0}}{g_{s0}}\right]+
\frac{T_3e^{3\phi_0}}{4g_{s0}}
\left[1-6\alpha e^{-2\phi_0}\left(H^2
+\frac{\dda}{a}\right)\right]\nn
&&+\sigma_0^2a^2\left[H^2
-6\alpha e^{-2\phi_0}\left(\frac{2\dda^2}{a^2}+\frac{2H\ddda}{a}
+\frac{9H^2\dda}{a}
+H^4\right)-18\alpha e^{2\phi_0}\left(\frac{\dda^2}{a^2}
+\frac{H\ddda}{a}
+\frac{3H^2\dda}{a}\right)\right]
\eea
Here we can see again that the contribution of those terms proportional
to $\sigma_0^2$ are supressed in late eras by at least one power of 
the scale factor. We do notice, however, that these contributions, 
to leading order in powers of the scale factor, \emph{i.e}. the terms of the form 
$\sigma_0^2 \, a^2 H^2$ in (\ref{rpi}),  do satisfy the equation of state of the DBI fluid (\ref{dbies}), if those terms were dominant.  
The constant terms $\frac{T_3e^{3\phi_0}}{4g_{s0}}$, on the other hand, 
play the role of the cosmological constant.
Thus we reproduce, as a further consistency check of our perturbative approach, the standard contributions to the density and the pressure
in the matter era, obtained previously, with our extra factors coming from the DBI part of the action
all being suppressed by the scale factor. This allows us to treat their effects as
subleading in the late eras of the Universe evolution, to which we shall restrict ourselves below. 

\subsection{The Perturbed Solution}
Let us now consider the perturbed metric; it reads
\bea\label{metrvar} g_{00} = -1+2\Psi~, \quad g_{ij} = a^2
(t)\left(1+2\Phi\right) \delta_{ij}~.  \eea
The perturbed vector field assumes the form
\be\label{timelike}
A_\mu = a(t) \left(- a(t)u_\mu + {\tilde A}_\mu \right)~,
\ee
where we used the background solution Eq.~(\ref{vectorcov}).

Caution should be exercised when we parametrise the vector field
perturbations, which should respect the time-like constraint for the
vector field Eq.~(\ref{constraint}). One should use the gauge
covariance associated with the general coordinate diffeomorphisms in
order to parametrise the temporal components of the vector
perturbations in terms of those of the metric~\cite{skordis}.  Below
we summarise briefly the situation and state the final result,
relevant to our purposes here. For details we refer the reader to
Ref.~\cite{skordis}. The reader should recall that, in contrast to the
TeVeS models, in our case the dilaton is constant and its
perturbations are ignored as sub-leading for reasons stated
previously. Moreover, the spatial components of the background vector
field are non vanishing, except when considering averages of
Eq.~(\ref{averages}) over populations of D-particles in the
foam. Nevertheless, these do not affect the general form of the vector
perturbations, which are of the same form as for TeVeS models.

The perturbation ${\tilde A}_\mu$ of the time-like vector field
$A_\mu$, Eq.~(\ref{timelike}), in a conformal metric background such
as
\be g_{00} = a^2(\eta)\ \ , \ \ g_{ij} = a^2(\eta)\ee
(where conformal time $\eta$ is related to cosmological time $t$ in
the standard way) can be decomposed as follows:
\begin{equation}\label{vectorpert}
\tilde{A}_\mu = - \Xi t_\mu + {\overline q}^\nu_{\,\,\mu} {\overline
  \nabla}_\nu \zeta + \beta_\mu~,
\end{equation}
where over-line denotes quantities pertinent to the conformal
background, $t_\mu $ is a time-like unit Killing vector tangent to the
geodesics, $t^\mu t_\mu = -1$, such that $t^\mu \beta_\mu = 0$, and
$\overline{q}^\mu_{\,\,\nu }= \delta^\mu_{\,\,\nu} + t^\mu \, t_\nu$
is a projector appropriate for the conformal metric~\cite{skordis},
such that ${\overline q}^\mu_{\,\,\nu} A^\nu = 0$, ${\overline
q}^\mu_{\,\,\alpha} {\overline q}^\alpha _{\,\,\nu} = {\overline
q}^\mu_{\,\,\nu} $. The fields $\zeta $ and $\Xi$ are scalar modes,
and $\beta_\mu$ are vector modes such that ${\overline q}^{\mu \nu}
{\overline \nabla}_\mu \beta_\nu = 0$. There are two independent
vector modes, as a result of the constraint Eq.~(\ref{constraint}) and
gauge invariance under diffeomorphisms.  The scalar mode $\zeta$
satisfies~\cite{skordis}
\begin{equation}\label{gf2}
{\overline \nabla}^2 \zeta = \vec{\overline \nabla} \cdot  \vec{{\tilde A}}~.
\end{equation}
There is gauge freedom in transforming the various fields under
general coordinate transformations, which results in simplifications
(upon gauge fixing) of the vector perturbations. To this end, let one
consider an infinitesimal diffeomorphism $\xi_\mu = \frac{1}{a}
{\widehat \xi}_\mu $, where $a$ is the scale factor of the conformal
metric. The vector ${\widehat \xi}_\mu $ can be decomposed as
${\widehat \xi}_\mu = - \xi t_\mu + {\overline q}^\nu_{\,\,\mu}
{\overline \nabla}_\nu \psi + \omega_\mu$, with $t^\mu \omega_\mu =
0$, which contains two scalar modes $\xi = t^\mu {\widehat \xi}_\mu $
and $\psi$, such that ${\overline \nabla}^2 \psi = {\overline
q}^{\mu\nu} {\overline \nabla}_\mu {\widehat \xi}_\nu$, and two vector
modes $\omega_\mu$ with the property ${\overline q}^{\mu\nu}
{\overline \nabla}_\mu {\omega}_\nu = 0$.  Under such diffeomorphisms,
the vector perturbation ${\tilde A}_\mu$ transforms as
\be\label{vectdiff} {\tilde A}_\mu' = {\tilde A}_\mu + \frac{1}{a}
\nabla_\mu \xi~; \ee
we recall that we ignore dilaton perturbations, which would in general
contribute.

From Eqs.~(\ref{vectorpert}) and (\ref{vectdiff}) we observe that the
scalar mode $\zeta $ transforms as
\be\label{scalardiff} \zeta ' =
\zeta + \frac{1}{a} \xi ~, \ee
the scalar $\Xi$ transforms as
\be\label{xidiff} \Xi ' = \Xi + \frac{1}{a} t^\mu {\overline
  \nabla}_\mu \xi~.  \ee 
while the vector modes $\beta_\mu $ remain gauge invariant.

Using the gauge freedom endowed in Eqs.~(\ref{scalardiff}),
(\ref{xidiff}) we may then gauge fix the perturbations of the vector
field, by making the gauge choice~\cite{skordis}
\begin{equation}\label{gf}
\Xi = -\Psi~, \quad \beta_\mu = 0~,
\end{equation}
where $\Psi $ is defined in the metric perturbation
Eq.~(\ref{metrvar}). Thus, the form of the (gauge fixed) vector
perturbation reads
\begin{equation}
{\tilde A}_\mu = \left[\frac{|T_3|e^{-\phi_0}}{g_{s0}}2\pi\alpha'\right]^{1/2}\left( \Psi, \vec{\tilde A} \right),
\end{equation}
with $\vec{\tilde A} $ satisfying Eq.~(\ref{gf2}), or equivalently
\be \vec{{\overline \nabla}} \zeta = \vec{{\tilde A}}~, \ee
setting constants to zero. A shift back to the Robertson-Walker metric
using a standard co-ordinate transformation leads to
\bea\label{a0} {\tilde A}_\mu =
  \left[\frac{|T_3|e^{-\phi_0}}{g_{s0}}2\pi\alpha'\right]^{1/2} \left(
  \frac{\Psi}{a(t)}, \vec{\tilde A} \right). \eea
Here the $\left[\frac{|T_3|e^{-\phi_0}}{g_{s0}}2\pi\alpha' \right]^{1/2}$ factor
is included from dimensional considerations.  The last perturbations
which need to be defined are the perturbations of the matter energy
momentum tensor. In
our system Eq.(\ref{Pfluid}) can be perturbed to find that \bea
\label{Emom}
\delta T^{{\rm m}0}_{~~~0} = -\delta\rho~, \quad \delta T^{{\rm m}i}_{~~~j} = \delta
P \delta^i_{~j}~.  \eea We can now begin to examine the perturbed
equations of motion, beginning with the graviton equation. In doing
so, it is useful to move into Fourier space, such that we get the
following conversion
\be \partial_i = ik_i~. \ee
This converts the partial differential equations into ordinary ones,
making them easier to manipulate.  The 00 component of the graviton
equation yields the following \bea T^{{\rm m}0}_{~~~0} &=&
-\left[\frac{1}{\kappa_0}+\frac{\alpha T_3
e^{\phi_0}}{g_{s0}}\right]\left[\frac{2k^2}{a}\Phi
+6H^2\Psi+6H\dot{\Phi}\right]\nn
&&+6\sigma_0^2a^2\Phi\left\{\frac{\dda}{a}+H^2-6\alpha
e^{-2\phi_0}\left[\frac{\dda^2}{a^2}+\frac{H\ddda}{a}+\frac{4H^2\dda}{a}-H^4+\frac{k^2}{3a^2}\left(2H^2-\frac{\dda}{a}\right)\right]\right\}\nn
&&+\Psi\biggl\{\frac{|T_3|(2\pi\alpha')e^{-\phi_0}}{g_{s0}}\frac{k^2}{2a^2}\left[1-6\alpha
e^{-2\phi_0}\left(H^2+\frac{\dda}{a}\right)\right]-6\sigma_0^2a^2\biggl[\frac{2\dda}{a}+3H^2-\frac{k^2(1+2/a^2)}{8a^2}\nn
&&~~~~~~-6\alpha
e^{-2\phi_0}\left(\frac{3\dda^2}{a^2}+\frac{3H\ddda}{a}+\frac{12H^2\dda}{a}-2H^4-\frac{k^2}{4}\left[\frac{\dda}{a}\left(\frac{1}{a^2}+\frac{7}{6}\right)+H^2\left(\frac{1}{a^2}-\frac{13}{6}\right)\right]\right)\biggr]\biggr\}\nn
&&-3\sigma_0^2a^2\dot{\Phi}\left\{H-6\alpha
e^{-2\phi_0}\left(\frac{11H\dda}{a}-3H^3+\frac{2H
k^2}{3a^2}\right)\right\}-3\sigma_0^2a^2\dot{\Psi}\left\{H-6\alpha
e^{-2\phi_0}\left(\frac{5H\dda}{a}+4H^3-\frac{H
k^2}{3a^2}\right)\right\}\nn &&+18\sigma_0^2a^2\alpha
e^{-2\phi_0}\ddot{\Phi}\left\{\frac{\dda}{a}+4H^2\right\}+18\sigma_0^2a^2\alpha
e^{-2\phi_0}\ddot{\Psi}+18\sigma_0^2a\da\alpha
e^{-2\phi_0}\dddot{\Phi}\nn
&&-\frac{k^2}{4a}\left(H\zeta+\dot{\zeta}\right) \left[1-6\alpha
e^{-2\phi_0}\left(H^2+\frac{\dda}{a}\right)\right]\left[\frac{2|T_3|(2\pi\alpha')e^{-\phi_0}}{g_{s0}}+3\sigma_0^2(2+a^2)\right]~,
\label{rho1}
\eea where $k^2 = k_i k^i $ is the amplitude of the spatial components
of the momentum scale. Similarly, the $ij$ component of the graviton
equation yields \bea T^{{\rm m}i}_{~~~j} &=&
-\left[\frac{1}{\kappa_0}+\frac{\alpha T_3 e^{\phi_0}}{g_{s0}}\right]
\left\{\frac{k^ik_j}{a^2}(\Psi-\Phi)+\delta^i_{~j}\left[\frac{k^2}{a^2}\Phi-\left(\frac{k^2}{a^2}-2H^2-\frac{4\dda}{a}\right)\Psi+2H\left(3\dot{\Phi}+\dot{\Psi}\right)+2\ddot{\Phi}\right]\right\}\nn
&&+2\sigma_0^2a^2\Phi\delta^i_{~j}\left\{H^2-6\alpha
e^{-2\phi_0}\left[\frac{2\dda^2}{a^2}+\frac{2H\ddda}{a}+\frac{6H^2\dda}{a}+H^4+\frac{7H^2k^2}{6a^2}\right]\right\}\nn
&&+2\sigma_0^2a^2\Psi\delta^i_{~j}\biggl\{\frac{3\dda}{a}+5H^2-\frac{k^2}{4a^4}+6\alpha
e^{-2\phi_0}\left(\frac{\dda^2}{a^2}+\frac{H\ddda}{a}-2H^4+\frac{k^2}{12a^4}\left[\frac{3\dda}{a}+H^2\left(3+10a^2\right)\right]\right)\biggr\}\nn
&&-12\sigma_0^2a^2\alpha
e^{-2\phi_0}\dot{\Phi}\delta^i_{~j}\left\{4H^3+\frac{H\dda}{a}\right\}+12\sigma_0^2a^2\alpha
e^{-2\phi_0}\dot{\Psi}\delta^i_{~j}\left\{3H^3+\frac{5H\dda}{a}\right\}\nn
&&-12\sigma_0^2\dot{a}^2\alpha
e^{-2\phi_0}\left(\ddot{\Phi}-\ddot{\Psi}\right)\delta^i_{~j}+6\sigma_0^2H^2\alpha
e^{-2\phi_0}k^i k_j(\Phi-\Psi)\nn
&&+\frac{k^2\delta^i_{~j}}{2a}\left(H\zeta+\dot{\zeta}\right)
\left[1-6\alpha e^{-2\phi_0}\left(H^2+\frac{\dda}{a}\right)\right]~.
\label{P1}
\eea In the above we have separated out those components which come
from the standard Einstein Hilbert part of the action from those terms
which appear due to the effect of the D-particles. The standard terms
are those which are proportional to $1/\kappa_0+\alpha T_3
e^{\phi_0}/g_{s0}$. At this point a very useful simplification can be
made. The $T^{{\rm m}i}_{~~~j}$ equation has two types of terms, those
proportional to $\delta^i_{~j}$ and those proportional to $k^ik_j$. By
contracting the equation with the following tensor operator \bea
\hat{k}_i\hat{k}^j-\frac{1}{3}\delta^j_{~i} \eea all the terms
proportional to $\delta^i_{~j}$, including those on the left hand side
of the equation are set to 0. Thus, after this operation on
Eq.~(\ref{P1}) we find that 
\bea
\label{Psi=Phi}
\Psi-\Phi = 0~.  \eea Thus, in our system, just as in the standard
gravitational case, $\Phi = \Psi$ is a valid solution which we shall
now be using throughout. As we shall show later on, the perturbations
of the recoil vector field $A_\mu$ do exhibit growing modes and thus
can participate in late-era structure formation, unlike the case of
TeVeS-like models in \cite{dodelson}, where there is a non-trivial
difference $\Psi - \Phi \ne 0$ that is considered as the main source
of the growing mode~\cite{feix}.

Upon using  Eqs.~(\ref{Psi=Phi}), (\ref{Emom}) and (\ref{bckgrndrho}), we obtain for the dimensionless over-density parameter $\delta \rho/\rho$ the result:
\bea
\frac{\delta\rho}{\rho}= \frac{a^3}{\Omega_m}\Biggl(&&\left[\frac{1}{\kappa_0}+\frac{\alpha T_3 e^{\phi_0}}{g_{s0}}\right]\left[2\Phi\left(\frac{k^2}{a^2}+3H^2\right)+6H\dot{\Phi}\right]\nn
&&-\Phi\biggl\{\frac{|T_3|(2\pi\alpha')e^{-\phi_0}}{g_{s0}}\frac{k^2}{2a^2}\left[1-6\alpha e^{-2\phi_0}\left(H^2+\frac{\dda}{a}\right)\right]-6\sigma_0^2a^2\biggl[\frac{\dda}{a}+2H-\frac{k^2(1+2/a^2)}{8a^2}\nn
&&~~~~~~-6\alpha e^{-2\phi_0}\left(\frac{2\dda^2}{a^2}+\frac{2H\ddda}{a}+\frac{8H^2\dda}{a}-H^4-\frac{k^2}{12a^2}\left[H^2\left(\frac{3}{a^2}+\frac{3}{2}\right)+\frac{\dda}{a}\left(\frac{3}{a^2}-\frac{1}{2}\right)\right]\right)\biggr]\biggr\}\nn
&&+6\sigma_0^2a^2\dot{\Phi}\left\{H-6\alpha e^{-\phi_0}\left(\frac{8H\dda}{a}-\frac{H^3}{2}+\frac{H k^2}{6a^2}\right)\right\}-18\sigma_0^2a^2\alpha e^{-2\phi_0}\ddot{\Phi}\left\{\frac{\dda}{a}+5H^2\right\}-18\sigma_0^2a\da\alpha e^{-2\phi_0}\dddot{\Phi}\nn
&&+\frac{k^2}{4a}\left(H\zeta+\dot{\zeta}\right)
\left[1-6\alpha e^{-2\phi_0}\left(H^2+\frac{\dda}{a}\right)\right]\left[\frac{2|T_3|(2\pi\alpha')e^{-\phi_0}}{g_{s0}}+3\sigma_0^2(2+a^2)\right]~~\Biggr)~.
\label{rho2}
\eea 

From this equation we can calculate the growth of density
perturbations, provided that the evolution of the metric and vector
perturbations are known. An important point to note is that the vector
perturbation terms, $\zeta$ and $\dot{\zeta}$ only appear due to the
$\langle \lambda (x) \rangle A_\mu A_\nu$ term of
$\mathcal{F}_{\mu\nu}$ in Eq.~(\ref{splitlameff}).  Thus the
constraint term is vital in coupling the vector perturbation to the
density perturbations.

\section{Growing Modes and large-Structure Formation due to D-particles}
\label{sec:growth} 

We now proceed to calculate the evolution of the vector field
perturbations. We notice that the remaining equations of motion
provide the constraints which allow the evolution of these
perturbations to be calculated. Under the perturbations of the metric
it is noticed that, since all the terms in the vector equation are
linear in the gauge field, any perturbations which appear from the
metric will always come with a $u_i$ term. Such terms, when averaged
over, are zero (as discussed previously). Thus, in our case metric
perturbations only appear in the vector equation through the presence
of $\Psi = \Phi$ in $\tilde{A}_0$ (\emph{cf}. Eq.~(\ref{a0})). The
perturbed vector equation reads
\bea\label{vecpert}
\ddot{\zeta}+b_1\dot{\zeta}+b_2\zeta = S[\Phi,C]~, \eea 
where 
\bea S[\Phi,C] &=&
\frac{\dot{\Phi}}{a}+\frac{\Phi}{a}\left(H+\frac{6\alpha
e^{-2\phi_0}\left[2H^3-\frac{\ddot{a}H}{a}-\frac{\dddot{a}}{a}\right]}
{1-6\alpha
e^{-2\phi_0}\left(H^2+\frac{\ddot{a}}{a}\right)}\right)~,\nn
b_1 &=& 3H+\frac{6\alpha
e^{-2\phi_0}\left[2H^3-\frac{\ddot{a}H}{a}-\frac{\dddot{a}}{a}\right]}{1-6\alpha
e^{-2\phi_0}\left(H^2+\frac{\ddot{a}}{a}\right)}~,\nn b_2 &=&
\frac{\ddot{a}}{a}+H^2+\frac{6\alpha
e^{-2\phi_0}H\left[2H^3-\frac{\ddot{a}H}{a}-\frac{\dddot{a}}{a}\right]-2\langle
\lambda (x) \rangle }{1-6\alpha
e^{-2\phi_0}\left(H^2+\frac{\ddot{a}}{a}\right)}~. \label{btwodef}\eea
where we remind the reader that $\langle \lambda (x) \rangle$ is the
background Lagrange multiplier, the form of which is given in
Eq.~(\ref{lambda}). In Eq.~(\ref{vecpert}) the signs of terms $b_1$
and $b_2$ play a crucial role in determining if the vector modes of
the perturbation enter a growing mode or a decaying one, a result
which is analogous to that found in Ref.~\cite{dodelson}. In general, growth will be
seen when the contribution from $b_2$ is negative.

\begin{figure}[t]
\begin{center}
\includegraphics[width=0.9\textwidth]{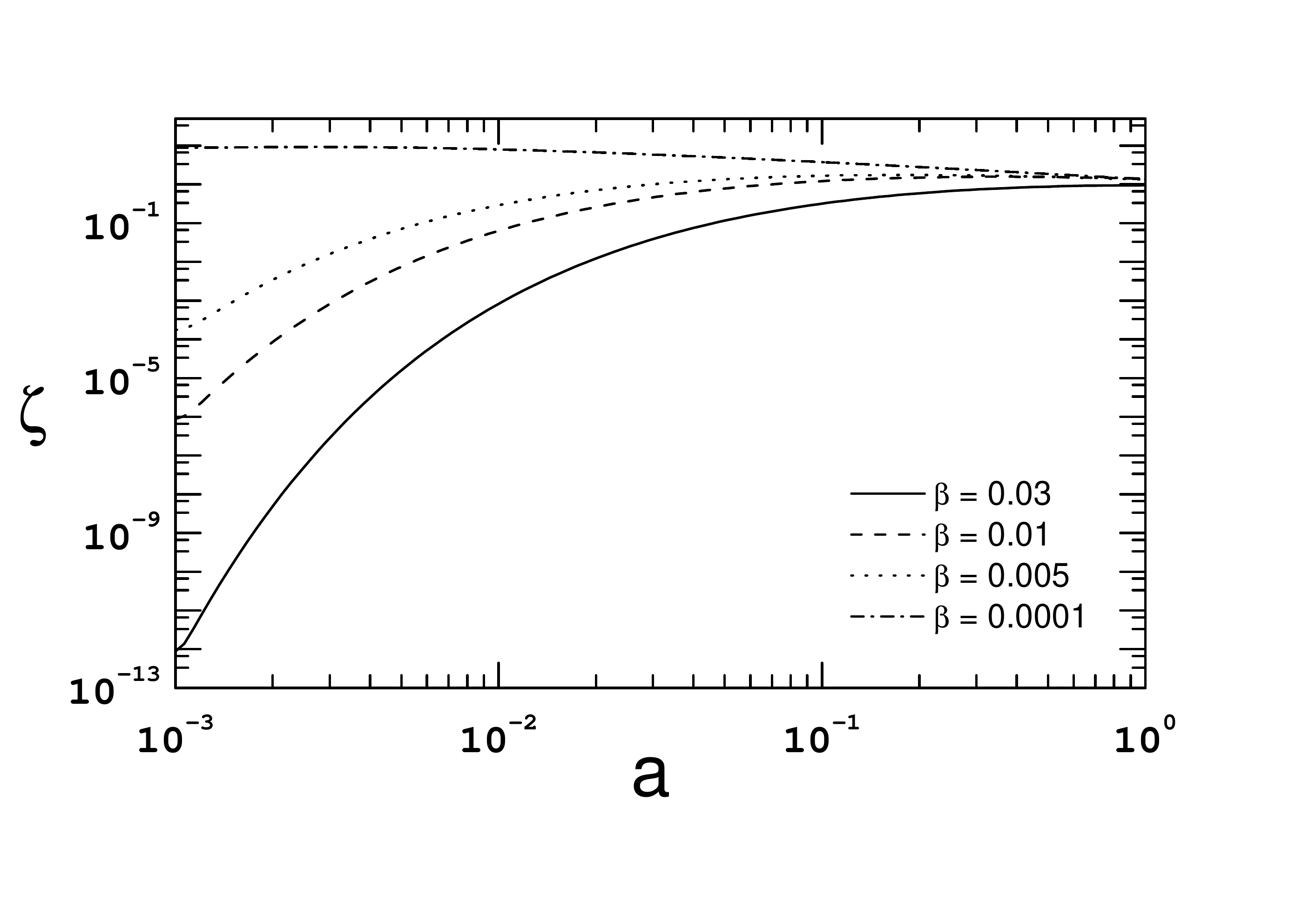}
\end{center}
\caption{Vector perturbation $\zeta$ as a function of the scale factor
$a(t)$ in the matter-dominated era, for different values of the
D-particle recoil velocity variance $\sigma_0^2 \propto \beta/a^3$. The string mass $M_s$ is assumed to be 10 TeV, $\phi_0=1$
and $k = 0.5~{\rm Mpc}^{-1}$.}
\label{fig:graph1}
\end{figure}

To this end, we first note that for large epochs of the Universe,
namely the radiation and matter dominated eras, it can be readily seen
that $\langle \lambda (x) \rangle > 0$.  In our analysis below we shall assume  values of $\phi_0 = \mathcal{O}(1)$, 
which may be phenomenologically desirable, since the string coupling $g_s = e^{\phi_0}$ determines the gauge couplings of the low-energy theory.
Looking at $b_2$ it can be
seen that in this case  the $\langle \lambda (x) \rangle$ term in crucial in
determining if $b_2$ will change sign~\footnote{We note for completeness that in models with large and negative values of $\phi_0$,  
the terms $\left[2H^3-\frac{\ddot{a}H}{a}-\frac{\dddot{a}}{a}\right]$ in $b_2$, although 
sub-leading at late eras of the Universe, since they contain higher orders of the scale factor
and its derivatives, nevertheless could have non-negligible contributions to $b_2$ due to the
denominator, which may be sufficiently small for sufficiently large and negative $\phi_0$, such that the over all
contribution of this term to $b_2$ is significant. In our case with $\phi_0=\mathcal{O}(1)$, though, such terms will not play any r\^ole in our analysis.}. Since 
$\langle \lambda (x) \rangle$ is proportional to  the variance of the recoil velocity (\emph{cf}. (\ref{lambda})), $\sigma_0^2$, we see that it is 
the magnitude of $\sigma_0^2$, which will determine if the vector perturbations will
enter a growing mode.  

Finally we examine the perturbed dilaton
equation. We get
\bea
\label{dilconst}
0&=&2\Phi\left\{12\sigma_0^2a^2\alpha\left[e^{-2\phi_0}\left(\frac{k^2H^2}{a^2}+6H^4+\frac{6H^2\dda}{a}\right)+6e^{2\phi_0}\left(\frac{H\ddda}{a}+\frac{\dda^2}{a^2}+\frac{3H^2\dda}{a}\right)\right]+\frac{\alpha
T_3e^{\phi_0}}{g_{s0}}\left[\frac{k^2}{a^2}+6H^2+\frac{6\dda}{a}\right]\right\}\nn
&&+6\dot{\Phi}\left\{12\alpha\sigma_0^2a^2\left[5H^3e^{-2\phi_0}+\frac{4H\dda
e^{2\phi_0}}{a}\right]+5H\frac{\alpha T_3
e^{\phi_0}}{g_{s0}}\right\}+6\ddot{\Phi}\left\{\frac{\alpha T_3
e^{\phi_0}}{g_{s0}}+12\alpha\sigma_0^2\da^2 e^{-2\phi_0}\right\}~.
\eea
We remind the reader that in our approach we assumed that any dilaton
couplings to matter have been suppressed, and this is the reason why
no matter contributions appear in the perturbations associated with
the dilaton constraint Eq.~(\ref{dilconst}). Here
we can see that the dilaton equation specifies the evolution of the
metric perturbations entirely. With the form of $\Phi$ then inserted in to
Eq.~(\ref{vecpert}), the form of $\zeta$ is found, and finally with
both $\zeta$ and $\Phi$ used in Eq.~(\ref{rho2}), $\delta\rho/\rho$ can
be calculated.

\begin{figure}[t]
\begin{center}
\includegraphics[width=0.9\textwidth]{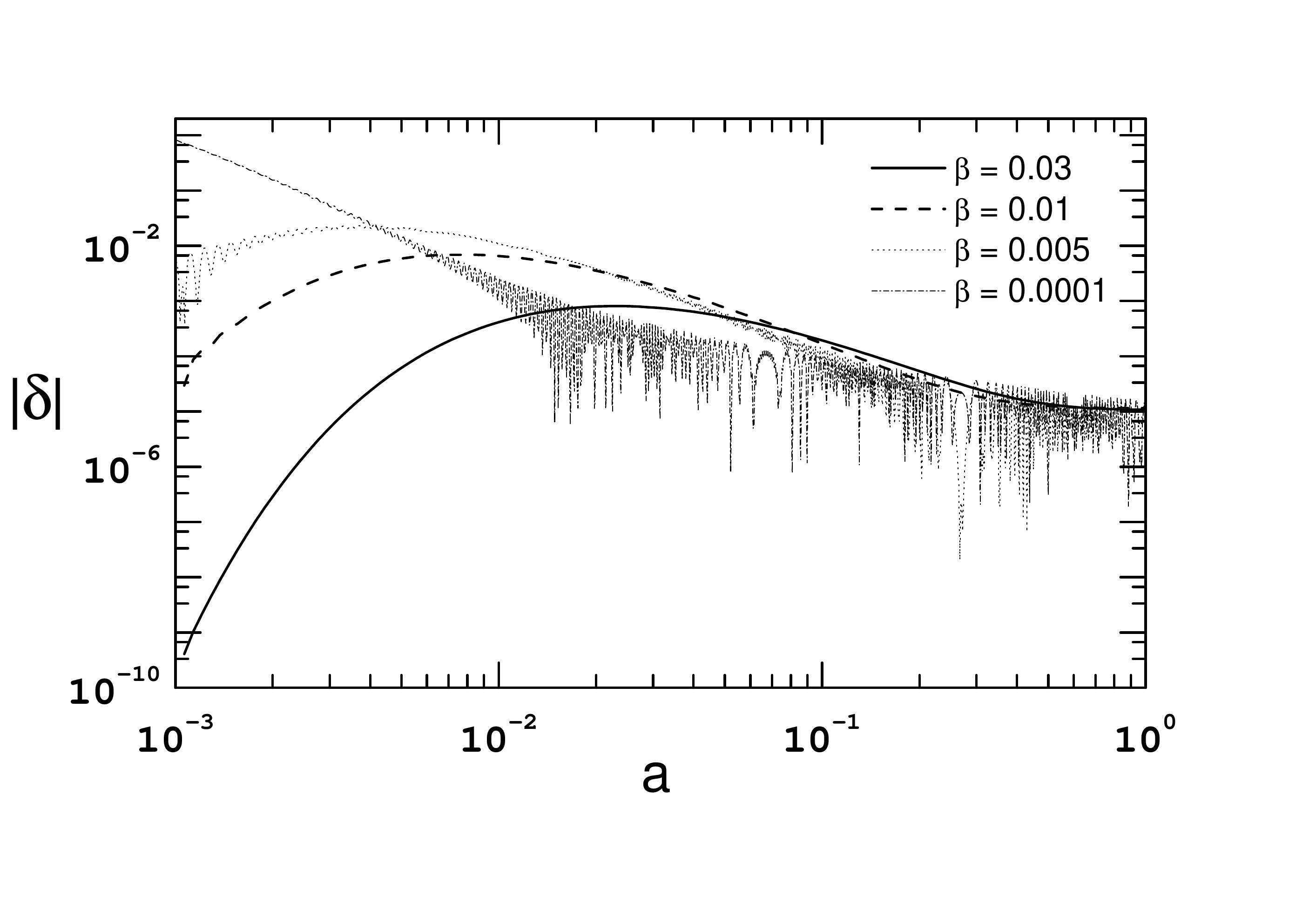}
\end{center}
\caption{Over-density parameter $|\delta|=\delta\rho/\rho$ as a
function of the scale factor $a(t)$ in the matter-dominated era, for
different values of the D-particle recoil velocity variance
$\sigma_0^2 \propto \beta/a^3$. The string mass $M_s$ is assumed to be 10 TeV, $\phi_0=1$
and $k = 0.5~{\rm Mpc}^{-1}$. $|\delta|$ is taken to be $10^{-5}$ today.}
\label{fig:graph2}
\end{figure}

The numerical results for the matter dominated era, that is a scale
factor $a \sim t^{2/3}$, are shown in Figs.~\ref{fig:graph1} and
\ref{fig:graph2}, which also shows the effect of altering the
D-particle recoil velocity variance,
(\emph{cf}. Eq.~(\ref{sigma_0dust})). A baryon only scenario was
considered for the numerical results as an extreme case, in order to
isolate the recoil-velocity effects of D-matter.  From the figures it
can be seen that the magnitude of the variance of the recoil velocity
of the D-particles plays the crucial role of allowing matter density
perturbations to grow sufficiently to allow for structure
formation. In particular, Fig.~\ref{fig:graph2} shows that when the
parameter $\beta$ is sufficiently small, $\beta \leq
\mathcal{O}(10^{-4})$, then $\delta\rho/\rho$ begins to show
oscillations and no longer exhibits growth. This is to be expected as
in the standard gravitational paradigm these same oscillations appear
in the absence of dark matter. The role of dark matter is in our
case being played by the coupling of the vector field perturbations to
$\delta\rho/\rho$. Thus for higher values of $\beta$ the vector
perturbations are strong enough that they can drive structure
formation in much the same way that dark matter does in the standard
case. Our point is not, however, to do away with dark matter
especially as dark matter candidates fall out naturally from the
string theory framework our approach relies upon. The point is rather
that there may be alternative methods of driving structure formation
which are in operation in conjunction with dark matter; the vector
perturbations of the approach presented here are one such candidate
which has now been shown to have the required characteristics to drive
structure formation. 

We can also note that the theory is largely insensitive to the string mass, 
with the
critical value of $\beta$ to induce the growing mode remaining approximately
of the same order over the range
 $10 ~{\rm TeV} < M_s < 10^{18}~ {\rm GeV}$. This can be readily seen
by looking once more at the expression for the coefficient $b_2$ appearing
in the perturbed vector equation, Eq.~(\ref{vecpert}). As we noted earlier,
only when $\phi_0$ is large and negative will those terms proportional to 
$\alpha = 1/M_s^2$ play a significant role. In the above numerical analysis
we took $\phi_0 = 1$, thus these terms have negligible effect. In the
background form of the Lagrange multiplier
 $\langle \lambda(x) \rangle$, given in Eq.~(\ref{lambda}),
 we see that the string mass dependence there is actually cancelled by
the string mass term appearing in the definition of $\sigma_0^2$ given
in Eq.~(\ref{sigma_0dust}). Thus $\langle \lambda(x) \rangle$ is
independent of the string mass and so the value of $\beta$ for which
the vector field will enter its growing mode is also independent of the
string mass. 

A further point of interest is the behaviour of the perturbed vector
equation in the radiation dominated era. In the radiation dominated
era the scale factor $a \sim t^{1/2}$ and as result,
\bea
H^2+\frac{\dda}{a} = 0\nn
2H^3-\frac{\ddot{a}H}{a}-\frac{\dddot{a}}{a} = 0
\eea
This can be used in the expression for the coefficient $b_2$ in 
 Eq.(\ref{btwodef}). $b_2$ now becomes,
\bea b_{2} = -2\langle \lambda (x) \rangle~. \eea Thus
in the radiation dominated era $b_2$ is always negative, since
$\langle \lambda(x) \rangle > 0$ (\emph{cf}. Eq.~(\ref{lambda})),
which as mentioned before is the condition for the growth of the
vector perturbations.  Thus in this era the vector perturbations are
always in a growing mode, irrespective of the specific density of the
defects, provided of course the latter is non zero and sufficiently
large so that the above formalism based on the existence of a
recoil-vector field is valid. Once again then the vector perturbations
show a behaviour similar to that of dark matter by being able to grow
during the radiation dominated era when baryon density perturbation
would not be able to cluster due to the radiation pressure. This
allows the seeding of the density perturbations which would drive
structure formation to begin early in the history of the universe.  As
the brane Universe enters the matter-dominated era, the growing mode
persists only above a critical density of defects, as we have
discussed previously (\emph{cf.}
Figs.~\ref{fig:graph1},\ref{fig:graph2}), which is an interesting
feature of the model.  Of course in the context of our string models,
in this era, other (conventional from a particle physics viewpoint)
candidates for dark matter, such as supersymmetric partners of mater
excitations or gravitinos do exist and contribute to the dark matter
spectra. However, in view of the assumption of sufficiently dense
populations of the defects on the brane, required for the
``\emph{medium'}' interpretation, the cosmological constraints of such
models, especially as far as collider searches for supersymmetry are
concerned~\cite{lmn}, are expected to be modified, depending on the
mass range of the D-particles~\cite{foam2}.

\section{Conclusions and Outlook}\label{sec:concl}

In this work, we have analysed some microscopic string theory models
of modified gravity arising in the low-energy limit of brane worlds
containing space-time point-like brane defects (D-particles).  Dense
media of such D-particles can exist at early eras of the Universe,
without the danger of over-closing the Universe, for specifically
stringy reasons.  Propagation of neutral matter, such as neutrinos, on
such backgrounds leads to effective gravitational theories containing,
in addition to the traditional graviton and dilaton fields of the
gravitational multiplet of the string, also \emph{vector} gauge
fields, describing the recoil of the brane defects during their
topologically non-trivial scattering with the string matter. The
vector fields are associated with the recoil velocities of the defects
and as such satisfy a given constraint Eq.~(\ref{constraint}). The
presence of a recoil velocity locally breaks Lorentz invariance, which
however is assumed to be restored on average over large populations of
D-particles, since the velocity expectation value vanishes, leaving
only the variances of the velocities to be non zero.

The effective Lagrangian describing the low-energy dynamics of this
model contains a Dirac-Born-Infeld type of lagrangian for the vector field,
coupled non-trivially to space-time curvature.  By solving the
associated equations of motion we have determined our background
configuration, over which perturbations were considered. We have also
demonstrated the consistency of the solutions with the conformal
invariance conditions of the associated stringy $\sigma$-model.  The
equation of state of the Dirac-Born-Infeld fluid of the vector field has
also been considered, and this result can be derived exactly to all
orders in $\alpha^\prime$.  The gravitational sector of the model,
unlike the vector field, cannot be studied exactly but only
perturbatively in powers of $\alpha^\prime$. In this work we
restricted ourselves to considering space-time curvature terms in the
effective action up to order $\mathcal{O}\big(\alpha^\prime\big)$,
which suffices for our low-energy considerations at late epochs of the
Universe.

By considering perturbations of the vector field, we have shown the
possibility of a growing mode in the matter era, for sufficiently
large values of the variance of the recoil velocities  of the D-particle.  The mode also
exists in the radiation era.  The constraint Eq.~(\ref{constraint}) of the
vector fields is essential in ensuring that the growing mode exists.
This allows the seeding of the density perturbations which would drive
large-structure formation to begin early in the history of the
universe.  This feature is shared by TeVeS models in alternative to
Dark matter scenarios, but the main difference of our model lies on
the fact that in our case the growing mode is independent of the
difference of the two gravitational potentials $\Psi - \Phi$ which vanishes here. This is consistent with the equations
of motion for the graviton and dilaton fields, and thus the associated
conformal invariance conditions of the stringy $\sigma$-model. In
addition, although in our theory, D-matter plays a role analogous to
dark matter as far as large-structure formation is concerned, this is
only one component, given that the underlying superstring inspired
models do involve additional components coming from the supersymmetric
partners of the standard model particle sector of the theory. In this
sense, the phenomenology of these models is different from
conventional low-energy string-inspired effective theories and can
depend crucially on the density of D-particle defects (and the
magnitude of their masses $M_{\rm s}/g_{\rm s}$) in the current era.
In this spirit, comparison of our class of models against
gravitational lensing data is essential in determining the amount of
dark matter present in the centre of galaxies, where the concentration
of D-matter is expected to be significant.

In our solutions in this article we assumed that the dilaton field was
constant.  Extensions of our model to incorporate time-dependent
dilaton fields are envisaged, although in such a case the
reconciliation of the model with the particle physics phenomenology
may be subtle, given that the exponential of the dilaton is connected
to the string coupling, and the latter to gauge couplings of the low
energy field theories coming from the string.  If we want standard
model physics to be reproduced today and the findings of the theory of
Big Bang Nucleosynthesis not to be disturbed, then one should apply stringent
constraints on the varying dilaton fields. Moreover, non constant
dilaton fields may have significant impact on the growth of galaxies,
and thus severe constraints from the relevant astrophysical data are
likely to be imposed, which can easily rule out most of such
cosmologies~\cite{growthmitsou}.

Finally, we would like to close by stressing the fact that the
detailed cosmology of such models is still in its infancy, but we
believe that the (rather toy) model we studied here exhibited several
interesting and non-trivial features that are worthy of further
detailed investigations in more realistic models and in very early
eras of the Universe. An interesting aspect we would like to study is
the role of the D-particles and the associated vector recoil field
in inflation. Such issues may be tackled by considering colliding
branes in our context (\emph{cf}. Fig.~\ref{fig:recoil}). Brane
collision would play the role of a cosmically catastrophic
event. During the collision, the concentration of massive bulk
D-particles that are trapped in the bulk space between the colliding
brane worlds can increase in such a way so that they can collapse to
form black holes. The latter would then evaporate, with the standard
model particles being trapped on the brane.  The gravitational
excitations of the string, on the other hand, are capable of escaping
in the bulk regions. The collision process can lead to inflationary
expansion on the brane universe~\cite{sakha}, while the black hole
evaporation can contribute to reheating of the inflationary universe
and subsequent graceful exit from the inflationary phase. The
particular role of the recoil vector field during the interaction of
string matter with the defects can be investigated at such very early
epochs of the Universe.  Although at present these considerations
appear to be mere speculations, nevertheless we believe that these
issues can be tackled at some detail within the framework of our
string-modified gravity. We hope to be able to report progress on some
of these topics in a future publication.

A final comment we wish to make concerns the density of D-particles on the brane world. Although cosmologically 
the later is left unconstrained, as we have already mentioned, given that the issue of overclosure of the brane Universe 
does not arise, nevertheless, the presence of a finite density medium of D-particles on the brane, with which neutral particles such as neutrinos and photons interact non trivially,
implies non-trivial ``optical'' properties for the brane, such as a  \emph{refractive} index. The latter will manifest itself as energy-dependent delays in the arrival times of photons emitted simultaneously from a high-energy astrophysical source, such as a Gamma Ray Burst or Active Galactic Nucleus. 
As discussed in \cite{refractive,mavroreview}, during each interaction of a high energy neutral particle (represented by a string) with a D-particle, there is a delay in the re-emission process of the particle due to capture which is linear in the incident particle energy $E$, $\delta t \sim E/M_s^2$, where $M_s$ is the string scale (we work in units of $c=1$).  The total delay turns out to be proportional to the number of D-particle defects per string length,  $\eta_{\ell_s}$, encountered by the propagating photon, times the distance travelled, $ \Delta t_{\rm total} \sim \frac{\eta_{\ell_s}}{M_s} E L$. This implies an effective suppression scale $M^{\rm eff} = M_s/\eta_{\rm \ell_s}$. The sensitivity of current observations of photons from sources at redshifts $z < 3$~\cite{mavroreview,refractive} to such linearly suppressed delays is such that  $M^{\rm eff}$  is of order of the Planck scale ($10^{19}$ GeV).  Since from particle physics experiments we know that $M_s \ge \mathcal{O}(10)$~TeV, we observe that an $\eta_{\rm \ell_s}$ as low as $10^{-15}$ can lead to observable effects from refractive index measurements for low string scales.  
In our models discussed here, we may easily imagine a depletion of defects in the bulk during epochs $z < 5$, so that their densities on the branes fall below such values, thereby 
avoiding  any constraint from the above refractive index measurements. 
If our defects on the brane (D-matter) are one or two orders of magnitude  more dense that the standard (conventional) dark matter, which is a quite natural situation, these optical constraints will still be negligible. 
Finally, we remark that the presence of D-matter may affect the peaks of the Cosmic Microwave Background Radiation, depending on the respective densities at that epoch, but this analysis requires a separate study, which goes beyond the purpose of our present article.

\section*{Acknowledgements} 

The work of N.E.M. was supported in part by the London Centre for
Terauniverse Studies (LCTS), using funding from the European Research
Council via the Advanced Investigator Grant 267352, and by STFC (UK)
under the research grant ST/J002798/1.


\end{document}